\documentclass[12pt]{elsarticle}      
\usepackage[english]{babel}
\usepackage{amsmath}
\usepackage[toc,page]{appendix}
\usepackage{amssymb}        
\usepackage{amsthm}         
\usepackage{mathbbol}
\usepackage{natbib}         
\usepackage{graphicx}       
\usepackage{subfigure}      
\usepackage{color}
\usepackage[export]{adjustbox}
\usepackage{algorithm}
\usepackage{algpseudocode}

\usepackage{tabularx, booktabs, multirow}
\usepackage{listings}
\usepackage{color}

\DeclareFixedFont{\ttb}{T1}{txtt}{bx}{n}{12} 
\DeclareFixedFont{\ttm}{T1}{txtt}{m}{n}{12}  

\usepackage{color}
\definecolor{deepblue}{rgb}{0,0,0.5}
\definecolor{deepred}{rgb}{0.6,0,0}
\definecolor{deepgreen}{rgb}{0,0.5,0}
\definecolor{red}{rgb}{1,0,0}
\usepackage{listings}

\newcommand\pythonstyle{\lstset{
language=Python,
basicstyle=\ttm\tiny,
otherkeywords={self},             
keywordstyle=\ttb\tiny\color{deepblue},
emph={MyClass,__init__},          
emphstyle=\ttb\tiny\color{deepred},    
stringstyle=\color{deepgreen},
frame=tb,                         
showstringspaces=false ,           %
commentstyle=\color{red},
breaklines=true
}}

\lstnewenvironment{python}[1][]
{
\pythonstyle
\lstset{#1}
}
{}

\newcommand\pythoninline[1]{{\pythonstyle\lstinline!#1!}}

\newcommand{\EPS}{\boldsymbol{\varepsilon}}
\newcommand{\SIG}{\boldsymbol{\sigma}}
\newcommand{\rd}{\mathrm{d}} 

\newcommand{\redcom}[1]{{ {#1}}}
\newcommand{\newredcom}[1]{{ {#1}}}

\begin{document}

\begin{frontmatter}

\title{An FFT based approach to account for elastic interactions in OkMC: Application to dislocation loops in iron} 

\author[urjc]{Rodrigo Santos-Güemes}
\ead{rodrigo.santos@urjc.es}

\author[labfus]{Christophe J. Ortiz}
\ead{christophe.ortiz@ciemat.es}

\author[upm,imdea]{Javier Segurado}\corref{cor1}
\ead{javier.segurado@upm.es}

\address[urjc]{Durability and Mechanical Integrity of Structural Materials Group (DIMME), Universidad Rey Juan Carlos, 28933, Móstoles, Madrid, Spain}

\address[labfus]{National Fusion Laboratory - CIEMAT. Avenida Complutense 40, 28040 Madrid, Spain}

\address[upm]{Departamento de Ciencia de Materiales, Universidad Politécnica de Madrid, C/ Profesor Aranguren s/n 28040 - Madrid, Spain}

\address[imdea]{IMDEA Materials Institute, 28906, Getafe, Madrid, Spain}


\begin{abstract}

Object kinetic Montecarlo (OkMC) is a fundamental tool for modeling defect evolution in volumes and times far beyond atomistic models. The elastic interaction between defects is classically considered using a dipolar approximation but this approach is limited to simple cases and can be inaccurate for large and close interacting defects. In this work a novel framework is proposed to include \emph{exact} elastic interactions between defects in OkMC valid for any type of defect and anisotropic media. In this method, the elastic interaction energy of a defect is computed by volume integration of its elastic strain multiplied by the stress created by all the other defects, being both fields obtained numerically using a FFT solver. The resulting interaction energies reproduce  analytical elastic solutions and show the limited accuracy of dipole approaches for close and large defects.

The OkMC framework proposed is used to simulate the evolution in space and time of self-interstitial atoms and dislocation loops in iron. It is found that including the anisotropy has a quantitative effect in the evolution of all the type of defects studied. Regarding dislocation loops, it is observed that using the \emph{exact} interaction energy result in higher interactions than using the dipole approximation for close loops.

\end{abstract}

\begin{keyword}
Object kinetic Monte Carlo, FFT homogenization, field dislocation mechanics, irradiation damage

\end{keyword}

\end{frontmatter}
\clearpage

\section{Introduction}
\label{sec:intro}

There are several scenarios with strong technological interest in which materials are exposed to severe irradiation conditions. These conditions are specially critical in materials for nuclear applications, for example in fission or in future fusion reactors \cite{Malerba10,Marian2017,Bjorkas07,Sand2014}. 
The harsh irradiation conditions produce a large amount of atomic displacements in the materials, which drives them out of equilibrium. As a result, it is observed that their macroscopic properties such as hardness~\cite{HajHam06}, ductile-to-brittle transition temperature~\cite{Schaeublin08}, conductivity~\cite{MorinMaita54, PearsonBardeen} or tensile strength~\cite{Wu05} are strongly affected. These effects point out the crucial interest of predicting the evolution of radiation damage in order to understand how material modifications evolve as function of irradiation parameters and/or to anticipate when the degradation of material properties will start affecting its performance. 

Considerable efforts have been devoted since decades to the experimental characterization and to the theoretical investigation of defects that form in materials under irradiation. Typical defects produced during the irradiation are point defects such as vacancies or self-interstitial atoms (SIA) \cite{Takaki1983}, their clusters thereof \cite{Bacon2000} and also larger defects such as dislocation loops \cite{Yao08, Prokhodtseva2013_2}.

Since the experimental characterization becomes very complex and experimental testing facilities allowing to perform tests under these conditions are extremely limited, modeling and simulation arises here as a fundamental tool to understand the evolution of defects. To this aim, different models or numerical approaches are available such as the Molecular Dynamics (MD)~
\cite{AIDHY201569,CHARTIER2019141}, kinetic Monte Carlo (kMC) and its different variants 
\cite{Soisson10, Terentyev11, Jimenez2016} or the Rate Theory (also sometimes called Cluster Dynamics by other authors) \cite{MARIAN201184,OrtizPichler, Jourdan14,LIU2017377,Kohnert2019}, based on coupled diffusion-reaction equations.

Each of these methods addresses the evolution of defects at different scales of time and space and thus has advantages and drawbacks. Since the degradation of macroscopic properties \textit{a priori} occurs over relatively long irradiation times, kMC method seems to be more appropriate than standard MD, which is limited to short times in the order of the ns. In particular, the subset of kMC denominated Object kMC (OkMC), which only follows the evolution of off-lattice defects allows reaching physical times close to those achieved experimentally. In this simulation framework, each object (defect or impurity) possesses a set of properties such as the migration energy or binding energy to other defects, which determines the frequency of the events it can undergo \cite{Domain04, Malerba06}. The OkMC approach follows the position of each individual defect and model their evolution considering discrete jumps through lattice positions. The model allows to account for spatial correlations and for complex interaction mechanisms between defects, which is an advantage over other models such as those based on the Rate Theory 
, which assumes a mean-field approximation. These features make thus the OkMC approach a valuable tool for the investigation of defect kinetics in irradiated materials in quasi-realistic conditions. 
In the absence of external forces and in an ideal perfect crystal lattice, defects can be considered as independent random walkers, i.e., their probability of jumps does not depend on the direction. However, in any realistic material, the host lattice is often locally distorted by the presence of external loads, dislocation lines, grain boundaries, small precipitates or other defects. This elastic distortion might bias the direction of the jumps depending on the elastic interaction energy of the defect when moving in different directions.

The strong effect of elastic fields in the evolution of defects or impurities has been studied in many different systems, including the migration of point defects as self-interstitial and vacancies near dislocations \cite{Bullough1970, Sivak2011}, the segregation of solutes at dislocation and grain boundaries in Fe-Cr alloys \cite{Kuksenko2013} or hydrogen atoms in W \cite{Mathew2020} and many other metals. In the case of the evolution of larger defects such as dislocation loops, the influence of elastic forces has been clearly evidenced experimentally. For instance, Dudarev \textit{et al.} observed using \textit{in situ} transmission electron microscope (TEM), how the migration of prismatic dislocation loops in irradiated iron becomes strongly correlated  due to their mutual long-range elastic interaction \cite{Dudarev2010}. A few years later, Mason \textit{et al.} \cite{Mason2014} observed in W under self-ion irradiation the presence of prismatic loops that are in principle highly mobile and should quickly recombine at the surface. The authors deduced that in fact, the loops are not isolated but interact through long-range elastic fields with other defects, which strongly affects their mobility. These experimental observations on loop interactions in metals were also supported by theoretical studies \cite{Dudarev2010, Dudarev2017,Mason2014}.

Hence, in the presence of heterogeneous elastic fields, the defects cannot be considered as random walkers and perform jumps with hopping probabilities biased by the spatially dependent elastic fields \cite{Hudson2005}. Although the continuum elastic description of defects is well developed \cite{Leibfried1978M, Mura87, Clouet2018}, the extension of OkMC simulations to account for these effects is relatively scarce and generally restricted to cases where analytic expressions for the elastic interaction energy are available. As example of these studies, Dudarev \textit{et al.} \cite{Dudarev2010} used a Langevin equation approach to predict the evolution of two interacting dislocation loops in Fe with their interaction energy being derived from theory of elasticity. Similarly, calculating the interaction energy between loops using isotropic expressions for elastic energy, Mason \textit{et al.} \cite{Mason2014} attempted to model the evolution of small dislocation loops that form in collision cascades in W and their elastic interaction using an OkMC model. Sivak et al \cite{Sivak2011} included elastic strain effects in a Kinetic Monte-Carlo simulation of self-point defect diffusion in bcc iron and vanadium using the anisotropic linear theory of elasticity. More recently, a few studies for computing the interaction energy can be found which still rely on a dipole approach, but in which some numerical technique is used to compute the external elastic field instead of using closed form expressions. As examples, in \cite{CARPENTIER2017323} a FFT solver was used for computing dislocation elastic fields in a anisotropic media and in \cite{Jourdan_2022} the use of a Fast Multipole approach was proposed to compute biharmonic strain fields as the ones created  by dislocation segments. 
It is also worth citing  the approach recently developed by McElfresh \textit{et al.} \cite{McElfresh2022} that considers localized stress effects by coupling an OkMC vacancy diffusion model with standard discrete dislocation dynamics. In this case, the elastic fields are calculated by the dislocation dynamics simulator and are updated as the dislocation microstructure evolves in time.

It can be seen that in most of the works mentioned above, analytical expressions are used to compute the elastic interaction energy in order to calculate the biased probabilities of jump of the defects in the different directions with two main considerations: (1) the expression of the elastic fields created by the interacting defects are given as a simple analytic closed expression (with some exceptions as \citep{CARPENTIER2017323,Jourdan_2022}) and (2) the interaction energy is based on the dipole tensor of the defect. Therefore, general cases considering the evolution of defects in complex nano-structures containing curved dislocation, phases with different  anisotropic elastic properties or free surfaces have not been studied yet. Moreover, for relatively large dislocation loops, the dipole tensor approximation for the interaction energy can be inaccurate when interacting defects are close. As a conclusion, a general OkMC framework is missing in which the defect evolution can be computed accurately for any defect distribution and in which interaction energies can be obtained without dipole expansion if needed. As we shall show throughout this work, this can be achieved by coupling the OkMC method with a boundary value solver of the continuum elasticity problem in the presence of defects. Thus, micromechanical FFT based methods, which have become a popular alternative to Finite Element in many cases \cite{Schneider2021,lucarini2021fft}, arise as an ideal numerical approach for this purpose.

The objective of this paper is twofold. On the one hand, a general novel computational method is proposed to include elastic interactions in OkMC simulations considering anisotropic elastic behavior and any defect distribution, curved dislocations and phases with different elastic properties. The approach does not rely on the dipolar approximation but instead is based in obtaining numerically the elastic fields of the defect map using a micromechanical FFT approach \cite{lucarini2021fft} combined with  defect eigenstrains or static field dislocation mechanics \cite{ACHARYA2001761,BERBENNI20144157}. On the other hand, the method developed will be used to analyze the evolution of self-interstitials and $<111>$ prismatic dislocation loops in Fe in the presence of other defects and immobile dislocations. These defects appear as consequence of the irradiation of steels under harsh environments such as those found in fission or in future fusion reactors and are responsible of the change of mechanical properties of steels \cite{Hardie13, Dethloff18}. Predicting their evolution and their interaction with the surrounding microstructure as accurately as possible is thus of crucial importance in the field of nuclear materials.

\section{Elastic energy interaction of defects}
\label{sec:FFT_energies}

Let be a crystalline (anisotropic) medium originally forming a perfect undeformed lattice. Let $d$ be a defect (or a collection of several defects) embedded in that medium, e.g. vacancy, interstitial, second phase or dislocation. The effect of the defects in the medium is the distortion of the lattice surrounding it. Some atomic positions away from a defect center this distortion can be described by continuum fields. The displacement field in the presence of a defect, $\mathbf{u}^d$, corresponds to the movement of lattice points from their original position. The total distortion $\mathbf{U}^d$ is defined as the gradient of the displacement field, and the strain $\boldsymbol{\varepsilon}^d$ is the symmetric part of the distortion, which is a compatible strain field,
\begin{equation}\label{eq:dist}
\mathbf{U}^d=\nabla \mathbf{u}^d \quad ; \quad \EPS^d = \mathbf{U}^{d,symm}
\end{equation}
The strain field $\EPS^d$ can be split into an inelastic part ($\EPS^{d,i}$), that is not proportional to the stress, and an elastic field, $\EPS^{d,e}$, which corresponds to the reversible part of the lattice deformation
\begin{equation}
 \EPS^d = \EPS^{d,e} + \EPS^{d,i}.
 \label{eq:el_inel}
\end{equation}
Unregarding its origin, the inelastic strain can be considered as an eigenstrain \cite{Eshelby1957} ($\EPS^{d,i}=\EPS^{EIG}$) in order to compute the associated elastic strain. The origin of this ineslastic strain depends on the type of defect and is \newredcom{concentrated in a very small region $\Omega_i$ (a few atomic positions) around the defect.} The elastic strain field, on the contrary, has a much larger range and decays relatively far from the defect center. The relative position between atoms in the inelastic region cannot be considered using elasticity and is governed by the atomic interactions and can be computed using atomisitic or DFT methods.

The elastic energy of the defect in absence of other defects and external loading is given by
\begin{equation}
E^{d}_0 = \int_{\Omega_d }\Psi(\EPS^{d,e}(\mathbf{x})) \rd \Omega,
\label{eq:Ed}
\end{equation}
where $\Omega_d$ is a volume surrounding the defect out of which the elastic strains caused by it can be neglected. Note that this region contains and is much larger than the region $\Omega_i$ where inelastic strains (or eignestrains) are defined. In Eq. \eqref{eq:Ed}, $\Psi$ is the elastic energy density at a point of the crystal, defined as
\begin{equation}
    \Psi(\EPS^{d,e})=\frac{1}{2}\mathbb{C}:\EPS^{d,e}:\EPS^{d,e} = \frac{1}{2}\SIG^{d}:\EPS^{d,e}.
   \label{eq:en_density}
\end{equation}
where $\mathbb{C}$ is the elastic stiffness tensor. Eq. \eqref{eq:Ed} corresponds only to the elastic contribution of the defect energy, but the total defect energy has also an inelastic contribution which in the case of dislocations is the core energy and that cannot be obtained with elasticity theory.

$\SIG^d$ is the stress field, defined as the derivative of the energy density with respect to the elastic strain originated by the defect.
\begin{equation}
\SIG^d = \frac{\partial \Psi}{\partial \EPS^{d,e}}=\mathbb{C}:\EPS^{d,e}.
\label{eq:sigEL}
\end{equation}
Note that the stress field is defined everywhere and is proportional to the elastic strain, which in the small region around the defects where inelastic strains are defined is different from the total strain. Therefore $\SIG^d$ containes both elastic and inelastic contributions.

Now consider the presence of other set of defects near the position of the original defect $\mathbf{X}^d$ inducing a new elastic strain field $\EPS^\mathrm{ext}(\mathbf{x})$. The elastic energy of this ensemble in the same region corresponds to 
\begin{equation}
E^\mathrm{ext} = \int_{\Omega_d} \Psi(\EPS^\mathrm{ext}(\mathbf{x})) \rd \Omega_d,
\label{eq:Eext}
\end{equation}
and the interaction energy can be defined as the difference between the energy of the full system and the sum of the energies of the two ensembles,
\begin{equation}
\Delta E^{d}  =  \int_{\Omega_d} \Psi(\EPS^\mathrm{ext}+\EPS^{d,e} ) \rd \Omega -E^{d}_0 - E^\mathrm{ext}.
\label{eq:interaction_energy_FFT}
\end{equation}
Using the definitions in Eqs. \eqref{eq:Ed}, \eqref{eq:en_density} and \eqref{eq:Eext}, the previous equation can be rewritten as
\begin{align}
\Delta E^{d}  =  \int_{\Omega_d} \frac{1}{2}\left( \SIG^\mathrm{ext}+\SIG^\mathrm{d}\right): \left( \EPS^\mathrm{ext}+\EPS^\mathrm{d,e}\right)\rd \Omega -E^d_0- E^\mathrm{ext} = \\ \int_{\Omega_d} \frac{1}{2}\left ( \SIG^\mathrm{ext}:\EPS^\mathrm{d,e} + \SIG^\mathrm{d}:\EPS^\mathrm{ext}\right)  \rd \Omega
\label{eq:interaction_energy2}
\end{align}
Finally, using the Betti's  reciprocity theorem, the interaction energy of a defect characterized by the elastic fields $\SIG^d,\EPS^{d,e}$ with some external elastic strain field $\EPS^\mathrm{ext}$ (with corresponding stress $\SIG^\mathrm{ext}$) can be obtained as
\begin{equation}
\Delta E^{d} =   \int_{\Omega_d}  \SIG^\mathrm{d}(\mathbf{x}):\EPS^\mathrm{ext}(\mathbf{x})  \rd \Omega =   \int_{\Omega_d}  \SIG^\mathrm{ext}(\mathbf{x}):\EPS^\mathrm{d,e}(\mathbf{x})  \rd \Omega.
\label{eq:interaction_energy3}
\end{equation}
The elastic interaction energy of the defect (Eq. \eqref{eq:interaction_energy3}) depends on its position $\mathbf{X}^d$ with respect the external field. Therefore, the energy landscape of the defect on different positions is modified by the elastic interaction energy and this change results in different jump probabilities depending on the defect final position.

\subsection{Dipole tensor approximation}
The elastic interaction of a defect in a external field can be approximated considering that external field is nearly constant in the region $\Omega_d$. In this case the integral of Eq. \eqref{eq:interaction_energy3} can be approximated as
\begin{equation}
\Delta E^{d} \approx   \left[ \int_{\Omega_d}  \SIG^\mathrm{d}(\mathbf{x})\rd \Omega \right] :\EPS^\mathrm{ext}(\mathbf{X}^d)  = - \mathbf{P}^d:\EPS^\mathrm{ext}(\mathbf{X}^d)
\label{eq:interaction_energy4}
\end{equation}
where $\mathbf{P}^d$ is the elastic dipole tensor of the defect \citep{Siems1968}, which characterizes the elastic interaction as function of its shape and atomic structure. The definition of the dipole tensor corresponds then to
\begin{equation}
\mathbf{P}^d = - \int_{\Omega_d}  \SIG^\mathrm{d}(\mathbf{x})\rd \Omega_d =- \int_{\Omega_d} \mathbb{C}:\EPS^{d,e}(\mathbf{x}) \mathrm{d}  \Omega = -\Omega_d \ \mathbb{C}:\overline{\EPS}^{d,e}
\label{eq:dipole_def}
\end{equation}
\noindent
where $\overline{\EPS}^{d,e}$ is the average elastic strain produced by the defect in the region $\Omega_d$. Considering the region $\Omega_i\in \Omega_d$ where inelastic strains of the defect ($\EPS^{d,i}$, usually identified with stress-free strain or eigenstrains) are different to zero, previous expression can be written as
\begin{equation}
\mathbf{P}^d = \Omega_i\ \mathbb{C}:\overline{\EPS}^{d,i}
\label{eq:dipole_def2}
\end{equation}
which is the usual definition of the dipole tensor as function of eigenstrain \cite{Mura87}. In appendix \ref{AppendixB} the equivalence of expressions \eqref{eq:dipole_def} and \eqref{eq:dipole_def2} is demonstrated.

The dipole tensor includes both elastic and inelastic effects since the stress induced by the defect near its position do not have only elastic origin. In the case of point defects the inelastic effects are fundamental and the dipole tensor can be obtained accurately without the limitations of linear elasticity by using discrete methods as DFT e.g. \cite{CARPENTIER2017323,Ma2019}. In the case of dislocations usually inelaslic contributions are neglected and closed expressions of the dipole tensor can be derived using elasticity \cite{Mura87,Clouet2018}. 


The dipole approach for computing elastic interactions, Eq. \eqref{eq:interaction_energy4}, is a very accurate approximation for point defects since strain fields are highly localized. For other defects as dislocation loops, a practical limitation is that simple closed expressions are only available in limited cases as small dislocation loops embedded in isotropic media. For other more general cases, e.g. loop interaction in anisotropic media although some analytical expressions can be found \cite{WANG1996293,OHSAWA20111071}, the formulas become much more complex. Moreover, even in the cases where these approximations can be found, their accuracy is very limited when defects are not sufficiently far away from each other, as it will be shown in next section. 
The alternative to the dipole approximations is the full evaluation of the interaction energy (Eq. \eqref{eq:interaction_energy3}). This evaluation can be done by double integration of Green's functions along the interacting defects, but this method requires numerical integration with a computational cost that grows with the square of the number of defects/dislocation segments.

For these reasons, there is a clear need for a robust and fast numerical approach for computing the elastic energy interactions of arbitrary defects in the framework of OkMC simulations. This numerical approach should allow the generation of strain maps for arbitrary defect arrangement, with a computational cost independent of the number of defects, in non-isotropic elastic media and under the eventual presence of material heterogeneities. Its integration in OkMC should also allow to consider both \emph{exact} energy interactions or dipole-tensor expansions depending on the conditions. 

\section{A FFT based framework for the numerical evaluation of interaction energies}
\label{sec:FFT_energies_2}

An spectral method is proposed to evaluate the elastic interaction energy of moving defects using Eq. \eqref{eq:interaction_energy3}. This approach is based on the efficient numerical resolution of the elastic fields generated by the interacting defects. The numerical method proposed for solving the elastic problem relies on the FFT algorithm and, in the case of dislocations, is based on the static field dislocation mechanics model. The benefits of this approach are the following:
\begin{itemize}
\item It is totally general and allows to compute elastic energy interactions between any defect type and size and with any external field.
\item Anisotropic elastic materials can be considered without any additional computational cost.
\item The effect of second phases, pores or free surfaces can be accounted for.
\item Periodicity is directly fulfilled by the use of a spectral representation of the fields, which is periodic by nature.
\end{itemize}

The procedure to obtain the interaction energy of a defect $d$ with an external strain field follows the Eshelby procedure \cite{Eshelby1957} and is summarized in the next points:
\begin{enumerate}
\item Compute the inelastic strain produced by the presence of defects, which can be considered as an eigenstrain for obtaining the mechanical equilibrium.
\item Obtain total strain field result of the inelastic strains generated by the defect solving an elastic problem in an infinite periodic medium. From the total strain, obtain the elastic strains and stress in the material.
\item Obtain the interaction energy using Eq. \eqref{eq:interaction_energy3}
\end{enumerate}
In addition to the interaction energy, this technique allows to obtain numerically the dipole tensor considering elasticity. The different steps of the procedure will be explained below.

\subsection{Inelastic strains produced by a defect}

The stress and strain produced by a defect can be obtained following the Eshelby procedure \cite{Eshelby1957}, being the defects considered as regions of the material which suddenly change their shape. \redcom{The transformation is characterized by a localized inelastic stress-free strain  which describes the change in shape and size of a region occupied by the base material when transformed in the absence of interaction with the surrounding matrix. Eshelby \cite{Eshelby1957} named this inelastic strain as eigenstrain ($\EPS^{d,i}=\EPS^{EIG}$ in Eq. \ref{eq:el_inel}) and the total strain produced by a defect can then be considered as the sum of the eigenstrain and elastic strain.}


In the case of point defects, as interstitials and vacancies, the eigenstrain is the ratio of the relaxation volume with respect the original atomic volume and is applied in an very small region around the deffect. The correct value of this eigenstrain should be obtained using atomistic or ab-initio techniques. In the case of dislocation loops, a displacement jump is introduced between the upper and lower parts of the planar area closed by the loop, $A_d$, with value and direction given by the Burgers vector $\mathbf{b}$. Following \cite{Eshelby1957} this eigenstrain can be defined as 
\begin{equation}
\boldsymbol{\varepsilon}^{EIG}(  \mathbf{x} ) =\left\{ \begin{array}{cc} 
\lim\limits_{e\rightarrow 0} \frac{1}{e} (\mathbf{b}\otimes\mathbf{n})^{sym} & \text{for} \quad  \mathbf{x} \in V_d \\ 
0 &  \text{elsewhere} 
\end{array}
 \right.
 \label{eq:eig_shearloop}
  \end{equation}
where $\mathbf{n}$ is the normal vector of the surface $A_d$ and $V_d$ is a volume of thickness $e$ formed by the extrusion of the surface along direction $\mathbf{n}$ symmetrically respect to the dislocation plane.

For other general dislocation arrangements not forming closed loops, eigenstrains cannot be computed as in the previous cases. Alternatively a general inelastic strain field, here called plastic strain due to its origin related to dislocations, can be computed using the static field dislocation mechanics approach \cite{ACHARYA2001761,BERBENNI20144157}. In a crystal containing an ensemble of dislocations, the distortion (Eq. \ref{eq:dist}) and can be split in its elastic ($\mathbf{U}^{d,e}$) and plastic parts ($\mathbf{U}^p$). The plastic contribution can be further decomposed into a compatible ($\mathbf{U}^{p,\|}$, curl free) and incompatible part ($\mathbf{U}^{p,\perp}$), following the Helmholtz decomposition:
\begin{eqnarray}
 \mathbf{U}^d= \mathbf{U}^{d,e}+\mathbf{U}^{p} \nonumber \\
    \mathbf{U}^p= \mathbf{U}^{p,\|}+\mathbf{U}^{p,\perp} 
\end{eqnarray}
From a macroscopic view point, the incompatible part of the plastic distortion is due to the presence of a non-zero net Burgers vector result of a density of Geometrically Necessary Dislocations (GNDs) at a given point. This incompatible part or GND density tensor is given by the Nye tensor $\boldsymbol{\alpha}$
\begin{equation}
\text{curl} ( \mathbf{U}^p)= \text{curl} ( \mathbf{U}^{p,\perp}) = -\boldsymbol{\alpha}
\label{curlUe}
\end{equation}
where $\boldsymbol{\alpha}$ can be defined as the net Burgers vector in a point.
In the original context, the Nye tensor definition assumes separation of scales, and the volume associated to a macroscopic point contains a finite ensemble of dislocations in a lower scale. At the microscale an alternative definition can be derived if, instead of lines, dislocations are represented by continuum fields \cite{ACHARYA2001761} represented by a Burgers vector density $\rho_b$ (Burgers vector per volume at a point) around the dislocation line. In this case, the  value of the Nye tensor at a point $x$ can be rigorously defined as
\begin{equation}
\boldsymbol{\alpha}(\mathbf{x}) = \int_{L} \rho_b(\mathbf{x-s}) \otimes \mathbf{t}(\mathbf{s}) \mathrm{d} L_s.
\label{eq:Nye2}
\end{equation}
Following Helmholtz decomposition, the incompatible part of the plastic distortion fulfills
$$\text{div} ( \mathbf{U}^{p,\perp}) =\mathbf{0.}$$
Applying the curl to Eq. (\ref{curlUe}) and operating, a Poisson equation is obtained which solution provides $\mathbf{U}^{p,\perp}$:
\begin{equation}
\text{div} ( \text{grad} (\mathbf{U}^{p,\perp})) = \text{curl} \ (\boldsymbol{\alpha})
\label{eq_EIG_real}
\end{equation}

This non-homogeneous partial differential equation can be solved using Green's functions. Moreover, if the solution is searched in a periodic box $L_x\times L_y \times L_z$, in which all the fields are periodic, then an explicit solution in Fourier space can be derived \cite{BERBENNI20144157}.  Note that periodicity does not have to be enforced but arises naturally from the use of a spectral representation of the fields.

Let $\hat{(\cdot)}$ be the Fourier transform of a function defined in real space (depending on $\mathbf{x}$) to Fourier space (depending on spatial frequency $\boldsymbol{\xi}$), then the incompatible plastic distortion in  Fourier space can be found directly 

\begin{equation} \label{eq_EIG_fourier}
\hat{U}^{p,\perp}_{ij}(\boldsymbol{\xi})=-i \frac{\xi_k}{\xi^2} \epsilon_{jkl} \hat{\alpha}_{il}(\boldsymbol{\xi}) \  \text{for} \ \boldsymbol{\xi}\neq \mathbf{0} 
\end{equation}
\noindent with $ \epsilon_{jkl} $ the permutation tensor.

To fully determine the plastic distortion caused by the presence of dislocations, the compatible part have to be added, follow \cite{Taupin2017,Kohnert2021} for more details. In a dynamic situation considering dislocation movement, the compatible plastic distortion corresponds to the area swept by the dislocations. For a static ensemble of dislocations,
this compatible plastic strain appears only in the presence of closed loops, and is considered constant throughout the volume. Therefore, this term can be added to zero frequency in Fourier space,

\begin{eqnarray}\label{eq:plastic_0}
\hat{U}^{p,\|}_{ij} (\mathbf{0})=\frac{1}{2V}(\mathbf{b}\otimes \mathbf{A}+\mathbf{A}\otimes \mathbf{b} )   
 & \text{for} \ \boldsymbol{\xi}= \mathbf{0}
\end{eqnarray}

The total plastic distortion is obtained solving Eqs. (\ref{eq_EIG_fourier},\ref{eq:plastic_0}) for the incompatible and compatible parts, respectively, and transforming the fields back to real space. The symmetric part of the plastic distortion is the plastic strain, which can be considered as if it were an eigenstrain (stress-free) field to compute the equilibrium. This allows to treat all these type of defects following the same procedure.

\begin{equation}
    \boldsymbol{\varepsilon}^{EIG}=\frac{1}{2}(\mathbf{U}^{p}+\mathbf{U}^{p,T})
\end{equation}

\subsection{Elastic problem with eigenstrains}

In the absence of an external strain, the problem to be solved is finding the stress field in equilibrium for a given distribution of eigenstrains ($\boldsymbol{\varepsilon}^{EIG}(\mathbf{x})$) produced by the presence of a set of defects. As in the previous section, the problem is defined in a periodic domain $L_x\times L_y \times L_z$, where all the fields are periodic and the linear momentum has to be conserved locally,
\begin{equation}
\text{div} \ \boldsymbol{\sigma}=\mathbf{0}.
\label{eq:equilibrium}
\end{equation}
Due to the periodicity, there are no external tractions to be considered in the energy. Moreover, if the periodic domain is sufficiently large with respect to the defect, the interaction of the defect with its periodic replicas is negligible and results are equivalent to an infinite body.

The elastic strain caused by the presence of defects is the difference between the total strain, $\boldsymbol{\varepsilon}$ and the inelastic strains (eigenstrains) (Eq. \ref{eq:el_inel}) and the stress is proportional to this elastic strain, Eq. \eqref{eq:sigEL}. In our problem $\boldsymbol{\varepsilon}^{EIG}$ is either known a priori, e.g. the eigenstrain of a dislocation loop given in Eq. \eqref{eq:eig_shearloop}, or is obtained solving the Fourier problem in Eqs. (\ref{eq_EIG_fourier},\ref{eq:plastic_0}).

Introducing the stress in Eq. \eqref{eq:equilibrium} and moving the eigenstrain to the right hand side, the problem can be written as 
$$
\text{div} ( \mathbb{C}:\boldsymbol{\varepsilon}^d(\mathbf{x}))=\text{div} \ (\mathbb{C}:\boldsymbol{\varepsilon}^{EIG}(\mathbf{x})).
$$
This equation is a non-homogeneous differential equation (the Poisson equation) in which the source term (right hand side) depends on the eigenstrain (input) and in which the solution is the total strain field. In the case of a homogeneous medium, this equation can be solved using the Green's functions resulting in 
\begin{equation}
\boldsymbol{\varepsilon}^d= \mathbb{\Gamma}_\mathbb{C} \ast (\mathbb{C}:\boldsymbol{\varepsilon}^{EIG})
\label{eq_EIG_elastic_real}
\end{equation}
where  $\mathbb{\Gamma}_\mathbb{C}$ is the fourth order tensor function defined from the directional derivatives of the Green's functions and $\ast$ denotes a convolution. 
The previous equation has a direct solution in Fourier space where convolution is replaced by multiplication
\begin{eqnarray}
\hat{\boldsymbol{\varepsilon}}^d(\boldsymbol{\xi})=\hat{\mathbb{\Gamma}}_\mathbb{C}(\boldsymbol{\xi}) : (\mathbb{C}:\hat{\boldsymbol{\varepsilon}}^{EIG}(\boldsymbol{\xi})).
\label{eq_EIG_elastic_fourier}
\end{eqnarray}
$\mathbb{\Gamma}_\mathbb{C}$ has a closed form expression as function of the Lamé coefficients of the material in the case of an isotropic medium \cite{Moulinec1994}, and for a general non-isotropic stiffness tensor $\mathbb{C}$  is given by
\begin{equation}\label{eq:basicgammacomplete}
\widehat{\Gamma}_{ijkl}(\boldsymbol{\xi})=\xi_l\xi_j\left[C_{ijkl}\xi_l\xi_j\right]^{-1}, \forall \boldsymbol{\xi} \neq 0
\end{equation}
where $\xi_i$ is the component $i$ of the frequency vector $\boldsymbol{\xi}$.

The result of problem in Eq. \eqref{eq_EIG_elastic_fourier} is the periodic strain field of the defect $\EPS^d$, and from it, elastic strain and stresses to equilibrate the eigenstrain can be obtained as
\begin{equation}
\EPS^{d,e}=\EPS^d-\EPS^{EIG} \ ; \ \SIG^d = \mathbf{C}:\EPS^{d,e}.
\label{eq:summaryEL}
\end{equation}

\subsection{Computing the interaction energy}

The elastic energy density map of a defect (Eq. \eqref{eq:en_density}) is finally obtained using the fields obtained as a result of the elastic problem, 
 $\SIG^d$ and $\EPS^{d,e}$ (Eq. \eqref{eq:summaryEL}). For an isolated defect $d$, the elastic energy can then be defined by integrating the energy density (Eq. \eqref{eq:en_density}) in the full periodic domain, and the result is independent of the defect position thanks to the periodicity.  The interaction energy with an external elastic field $\EPS^{ext}$ caused by other defect or a far field strain is given by Eq. \eqref{eq:interaction_energy3}.  In this case, the position of the defect $\mathbf{X}^d$ with respect to the external strain determines the interaction energy. 
 
 As a defect moves, the elastic field caused by this defect translates with it. In order to compute the elastic interaction energy of a moving defect with an external field is sufficient to compute once the elastic strain field on a reference position $\mathbf{X}^0$, $\EPS^{d_0,e}(\mathbf{x})$ and translate this field considering the current position of the defect $\mathbf{X}^{d,e}$ and then evaluate Eq.  \eqref{eq:interaction_energy3}. The new field for the defect in position $\mathbf{X}^d$ is then given by
$$ \EPS^{d,e}(\mathbf{X}^d;\mathbf{x})=\EPS^{d_0,e}(\mathbf{x}-\mathbf{X}^d)$$
where $\mathbf{X}^d$ refers to the defect center position, and $\mathbf{x}$ to a position in the space.

An efficient way of performing this translation consists in using the shift theorem, a property of the Fourier transform,

\begin{equation}
\widehat{\EPS}^{d,e}(\mathbf{X}^d;\mathbf{x})=\widehat{\EPS}^{d_0,e}(\mathbf{x-X}^d)=e^{-i\boldsymbol{\xi}\cdot\mathbf{X^d}} \ \widehat{\EPS}^{d_0,e}(\mathbf{x}).
\label{eq:shift}
\end{equation}
Following this approach, the Fourier transform of the elastic strain field of the defect on a reference position $\widehat{\EPS}^{d_0,e}(\mathbf{x})$, is computed and stored once, and then for each new position $\mathbf{X}^d$ is just obtained by performing the inverse Fourier transform of Eq. \eqref{eq:shift}

\subsection{Numerical resolution using the FFT}

The Eqs. \eqref{eq_EIG_real} and Eq. \eqref{eq_EIG_elastic_real} for a homogeneous medium have an explicit solution in Fourier space,  Eq. \eqref{eq_EIG_fourier} and \eqref{eq_EIG_elastic_fourier}, and can be solved in that space by direct substitution without iterating. For solving the equations in a discrete form, the cubic domain is discretized first in $N_1\cdot N_2\cdot N_3$ equal sized voxels, with positions
\begin{equation}
x_{k}=\frac{L_k}{N_k}(\frac{1}{2}+n_k)
\end{equation}
and the functions involved in the equations are discretized by their value at the center of each voxel (e.g.
$\boldsymbol{\varepsilon}(\mathbf{x}_k)$), with $n_k \in [0,N_k-1]$. The corresponding discrete frequencies in Fourier space are $\boldsymbol{\xi}_k=n\frac{2\pi}{L_i}$ with $n$ given by
\begin{equation}\label{frequencies}
n=\left\{
\begin{array}{c} 
m-\frac{N_k-1}{2} \quad \text{for N}_k \ \text{odd}\\
m-\frac{N_k}{2}  \quad \text{for N}_k \ \text{even}
\end{array} 
\right. \ m=0, \dots, N_k-1 
\end{equation}

Then, the right hand sides of Eq. 
\eqref{eq_EIG_real} and Eq. \eqref{eq_EIG_elastic_real}  are transformed to Fourier space through the FFT algorithm. 
The discrete values of the frequencies are substituted on the operators in Eqs.  \eqref{eq_EIG_fourier}, \eqref{eq_EIG_elastic_fourier} and \eqref{eq:basicgammacomplete}  to obtain the eigenvalues corresponding to the prescribed Nye tensor density and the elastic strain fields, respectively. The final step is the transformation of the resulting fields back to real space using the inverse FFT algorithm. All these equations have been implemented in the FFT-homogenization code FFTMAD \cite{lucarini2021fft}. 

The proposed numerical framework is inherently more computationally intensive than classical analytical approaches, but it significantly broadens the range of cases to which it can be applied. Unlike classical methods constrained to simple scenarios with readily available analytical expressions or dipole tensors, this framework offers greater flexibility. Moreover, in analytical approximations, the computation of interaction energies scales with $n^2$, where $n$ represents the number of defects. In contrast, the new procedure scales linearly with $n$ which is particularly advantageous in simulations involving a large number of defects.

\section{Validation of the method and comparison with dipole approximations}
\label{sec:validation}

To analyze the validity of the FFT method described in previous section
the interaction energies of different defects 
will be obtained with the full integration method proposed on both isotropic and anisotropic matrix and compared with the exact solution when available and the dipole approximation.

The material used for the comparisons in all the examples will be iron and the properties are taken form \cite{Ma2019} and summarized in table \ref{tab:Fe}. In this reference the three elastic properties characteristic of cubic symmetry are given. In order to obtain two isotropic material constants from the three parameters of the real crystal different approaches can be followed as for example Voigt (iso-strain) and Reuss (iso-stress) approaches. In this study, we opt to compute the isotropic elastic parameters such that they preserve the response of the crystal for uniaxial cases in the orthotropic directions. To this aim, $C_{11}$ and $C_{22}$ are kept and Zener ratio of 1 is imposed $Z=1=C_{44}/(C_{11}-C_{12})/2$ so the shear modulus $C_{44}$ is then redefined as $C_{44}=(C_{11}-C_{12})/2$. \
\begin{table}[htp]

\begin{center}

\begin{tabular}{|c|c|c|c|c|}
\hline
C$_{11}$ (GPa) & C$_{12}$ (GPa) & C$_{44}$ (GPa) & E$_{iso}^{(1)}$  (GPa)& $\nu_{iso}^{(1)}$   \\ 
\hline
289.34  & 152.34 &107.43 & 184.25 & 0.344  \\
 \hline
\end{tabular}

\begin{tabular}{|c|c|c|}
\hline
 $a_0$ (\r{A})  & $V_{at}$ (\r{A}$^3$)  & $\Delta V^{(2)}$ (\r{A}$^3$) \\
\hline
2.83 & 11.34 & 1.620\\
\hline
\end{tabular}

\begin{tabular}{|c|c|c|c|c|c|}
\hline
P$_{11}$(eV) & P$_{22}$ (eV) & P$_{33}$ (eV) & P$_{12}$ (eV) & P$_{13}$ (eV) & P$_{23}$ (eV) \\
\hline
25.832  &21.143& 21.150 &0.000 & 0.000 & 5.122 \\
\hline
\end{tabular}
\end{center}
\caption{Elastic properties of iron \cite{Ma2019}. $^{(1)}$ Isotropic properties are obtained forcing $C_{44}=(C_{11}-C_{12})/2$. $^{(2)}$ The relaxation volume of the defect and the Dipole tensor components are  from a $<110>$ dumbbell self-interstitial.}
\label{tab:Fe}
\end{table}

For the FFT simulations, the $x,y$ and $z$ directions correspond to $\mathbf{x}=[1,1,1]/\sqrt{3}$,  $\mathbf{y}=[-1,-1,2]/\sqrt{6}$ and  $\mathbf{z}=[1,-1,0]/\sqrt{2}$ crystallographic directions. All simulations used the same discretization of $129^3$ voxels. The simulation box is a cube of 500$a_0$ on each direction. The Burgers vector of the dislocation, both for an infinite straight dislocation and for a dislocation loop, is spread to the Fourier points inside the dislocation core as a function of the distance. The magnitude of the dislocation core in iron is around 6.7 \r{A} ( = 2.7b ) according to MD simulations \citep{Osetsky2000}. Considering the size of the domain and the discretization employed in this work, the dislocation core is on the order of the voxel size, resulting in a spreading of the dislocation core to the 8 Fourier points surrounding each subsegment of the dislocation.

\subsection{Interaction energy of a self-interstitial with a straight dislocation}
\label{subsec:Eint_SIA_disloc}

The elastic interaction of a self-interstitial atom with a straight dislocation in Fe is analyzed here. The interaction is computed with the dipole approach (Eq. \eqref{eq:interaction_energy4}) because the strain field of a SIA is localized in a very small region and can be considered as a point from the continuum view point without any error. The SIA considered here is the $<110>$ dumbbell, i.e. the most stable one in Fe, and its dipole tensor is given in table \ref{tab:Fe}.

In the present case, the external elastic field is due to a straight edge dislocation, having its line in the direction $\mathbf{y}$, a Burgers vector with a value of $\sqrt{3}/2 a_0$ oriented in the direction $\mathbf{x}$, and passing through the center of the cell. Three different approaches are used to compute the interaction energy between the straight dislocation and the SIA; in (a) the straight dislocation field is computed using the analytical expressions of Volterra \cite{Volterra1907}, in (b) the field of the straight dislocation is obtained using the FFT approach and solving the field dislocation mechanics problem in an isotropic medium and in (c) the field is also obtained numerically, but considering the anisotropy of Fe.

\begin{figure}[ht]
\centering
\includegraphics[width=0.60\textwidth]{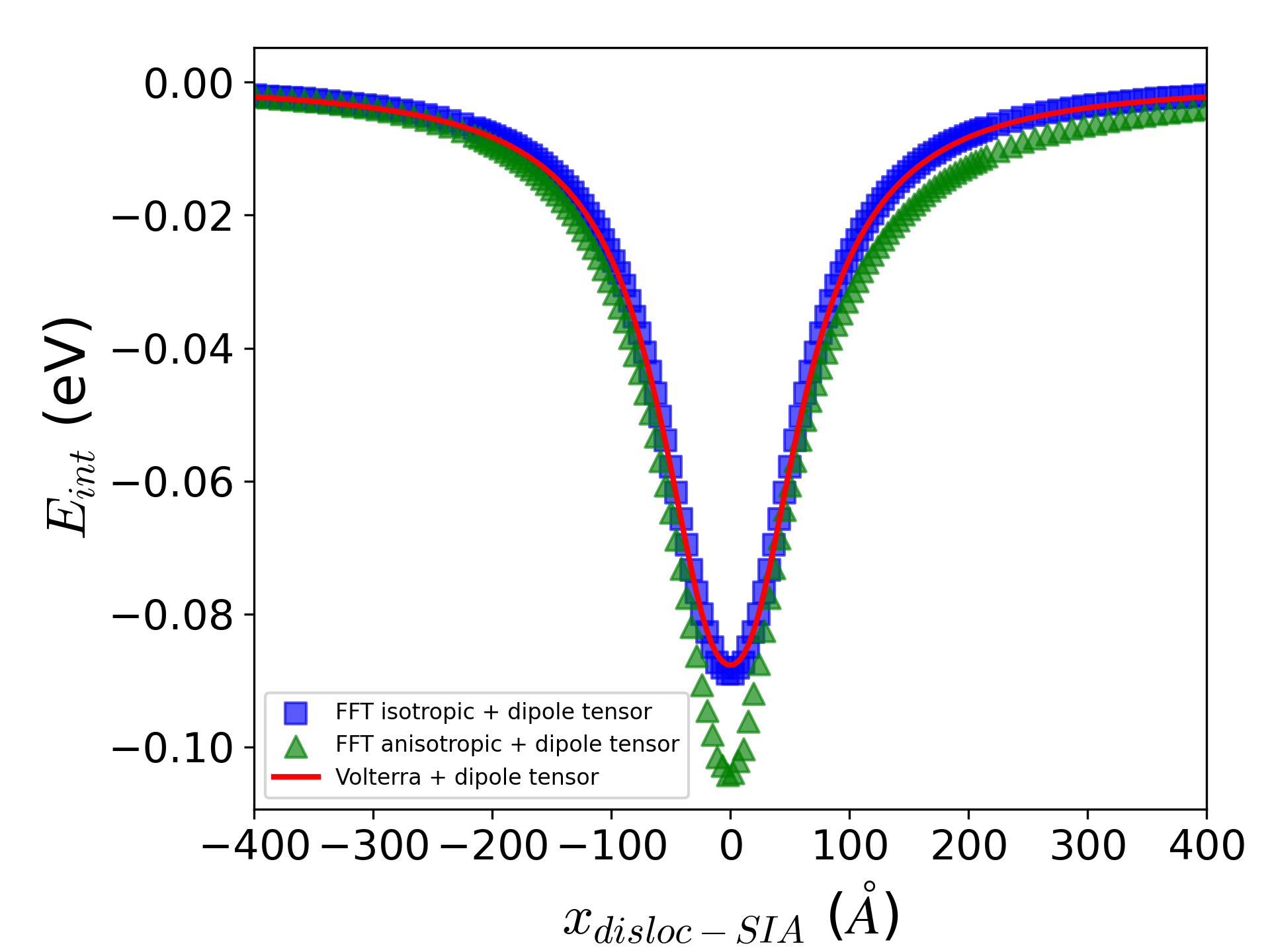}
 \caption{Elastic interaction energy of a straight dislocation and a SIA in Fe as a function of their relative distance along the $x$ axis using the analytical expressions of Volterra for the straight dislocation field (red line) or using the field obtained numerically by solving the dislocation mechanics problem in FFT for both isotropic (blue squares) and anisotropic (green triangles) matrix.} 
 \label{fig:Eint_disloc-SIA}
\end{figure}

For the FFT calculations, the same discretization, box orientation and size as in the previous example are used. The SIA is located on different locations of the box, with its center fixed in $z$ direction, $z_{SIA} = z_{disloc}$ + 50 {\AA} and varying its $x$ position along the box. The configuration is symmetric in the $y$ direction so this position is not relevant. The results of the calculations made with the analytical solution and those obtained with the FFT method are shown in Fig. \ref{fig:Eint_disloc-SIA}.

\subsection{Interaction energy of a dislocation loop and a straight dislocation}
\label{subsec:Eint_DL_disloc}

\begin{figure}[ht]
\centering
\includegraphics[width=0.47\textwidth]{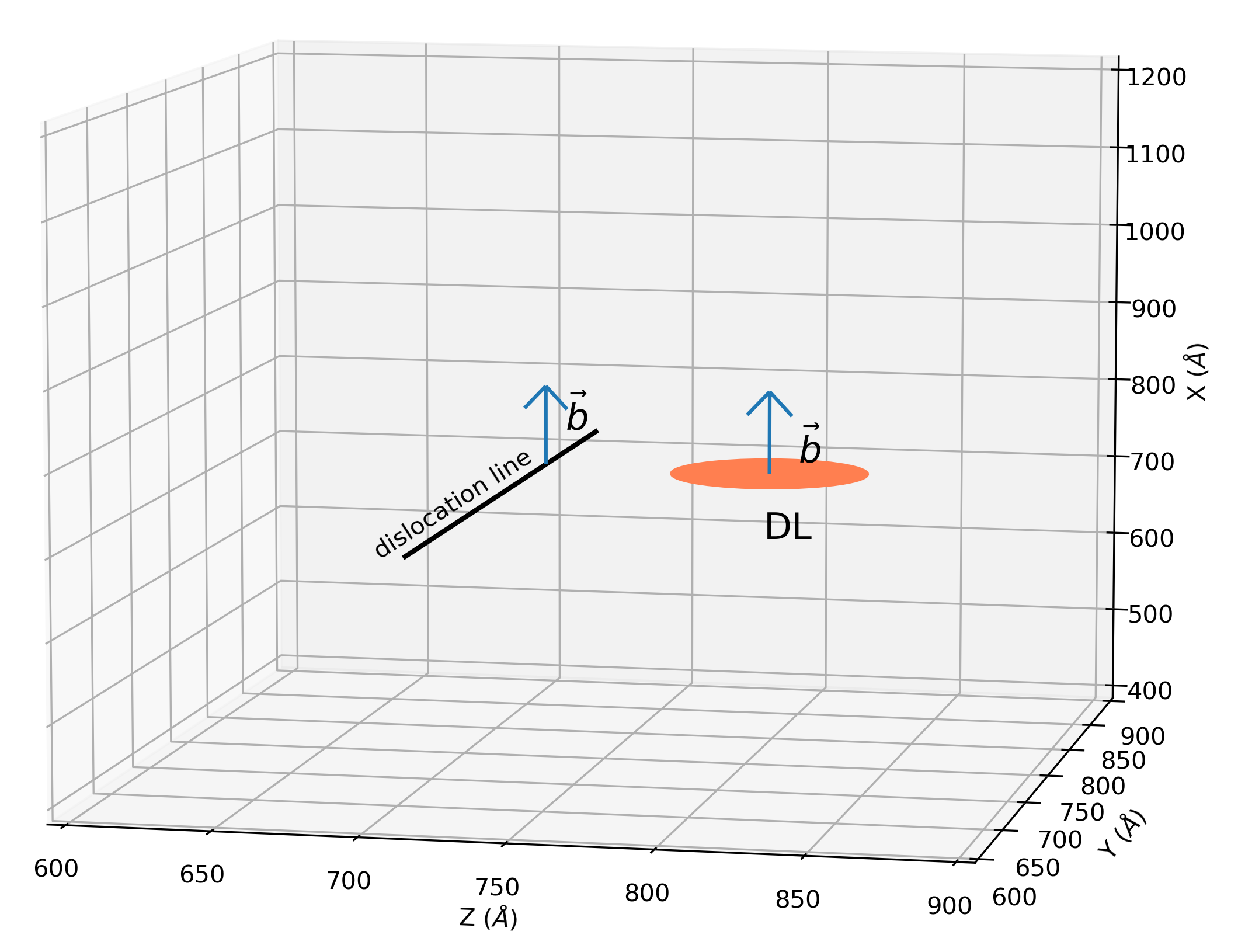}
\includegraphics[width=0.47\textwidth]{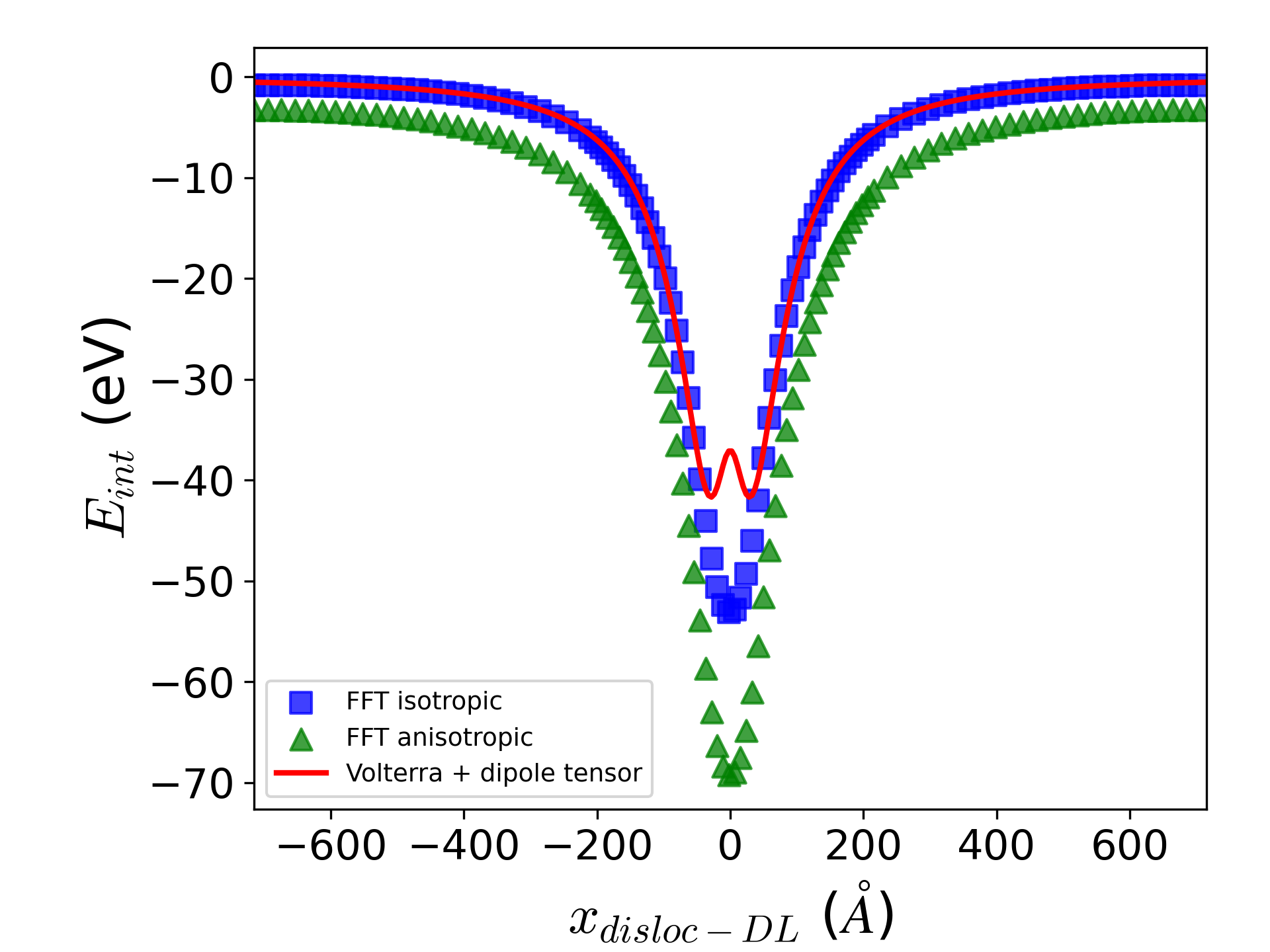}
 \caption{Left: Configuration of the straight dislocation and the prismatic loop in the simulation box. Right: Elastic interaction energy of the defects in iron as a function of their relative distance along the $x$ axis using the dipole approximation and analytical expressions of Volterra for the straight dislocation field (red line), or using a full numerical evaluation of the interaction energy based on the FFT approach for both isotropic (blue squares) and anisotropic (green triangles) matrix.} 
 \label{fig:Eint_disloc-DL}
\end{figure}

A system formed by a straight dislocation and a prismatic loop is now considered. The straight dislocation is the same as the one described in previous subsection. The dislocation loop contains 800 SIAs and has a Burgers vector $1/2\langle 111 \rangle$ 
and plane normal oriented in the same direction. Its dipole tensor is obtained from the Eshelby approach as
\begin{equation}
\mathbf{P}^d =\frac{1}{2} \ \mathbb{C}: (\mathbf{b}\otimes \mathbf{A}+\mathbf{A}\otimes \mathbf{b} )
\label{eq:dipole_DL}
\end{equation}
where $\mathbf{A}$ is the oriented area of the loop, which corresponds to a radius of approximately 34 {\AA}. This type of loops migrates in Fe by 1D glide along the direction of their Burgers vector, here the $\mathbf{x}$ direction. Contrary to the SIA, the dislocation loops are not point defects and the dipole approximation for obtaining their interaction energy (Eq. \eqref{eq:interaction_energy4}) with a straight dislocation can be inaccurate, in particular for short distances. Three different methods are considered here to compute the interaction energy; (a) fully analytical calculation using the dipole approximation (with dipole tensor given by Eq. \eqref{eq:dipole_DL}) and the strain field calculated with Volterra equations, (b) numerical calculations in which the field of both defects are computed using the FFT method assuming an isotropic matrix and (c) numerical calculation with the FFT method taking into account the anisotropic elasticity. The DL center was placed at a distance $z_{DL} = z_{disloc}$ + 50 {\AA} on the $z$ axis and its position was varied along the $x$ axis, i.e., along the direction where it can perform jumps. A sketch of the dislocation and the loop as well as the orientation of their Burgers vectors are represented on the left side of Fig. \ref{fig:Eint_disloc-DL}. The interaction energy between the dislocation and the loop obtained using the analytical approach as well as those obtained using the FFT method for isotropic and anisotropic media are shown on the right side of Fig. \ref{fig:Eint_disloc-DL} as a function of their relative distance along the $x$ axis.

\subsection{Interaction energy of two dislocation loops}
\label{subsec:Eint_DL_DL}

Finally, a system formed by two prismatic loops is considered. As before, both loops have the same radius of approximately 33 {\AA} and have Burgers vector and plane normal oriented in the $\mathbf{x}$ direction, which is also the climbing direction for their movement.

\begin{figure}[ht]
\centering
\includegraphics[width=0.45\textwidth]{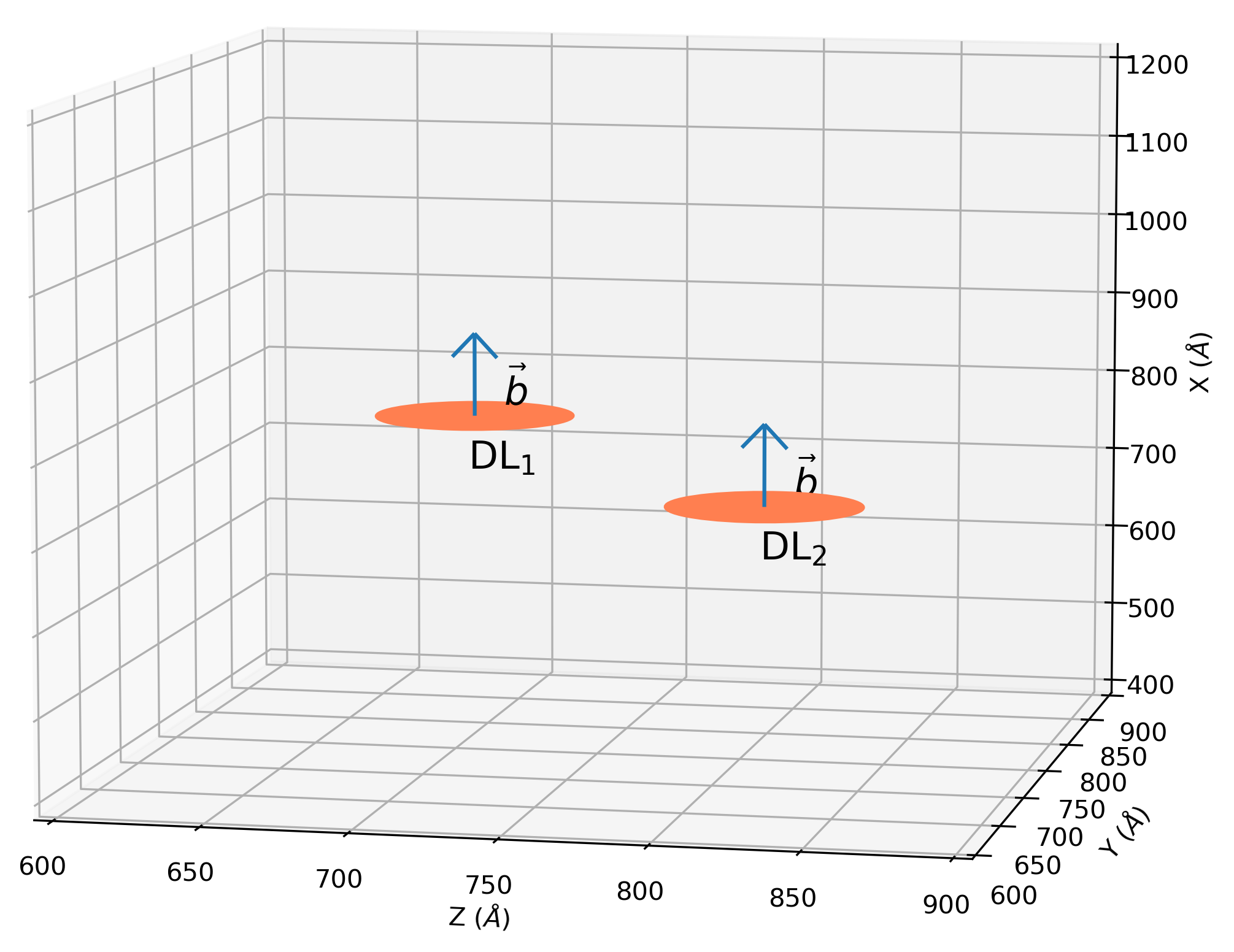}
\includegraphics[width=0.47\textwidth]{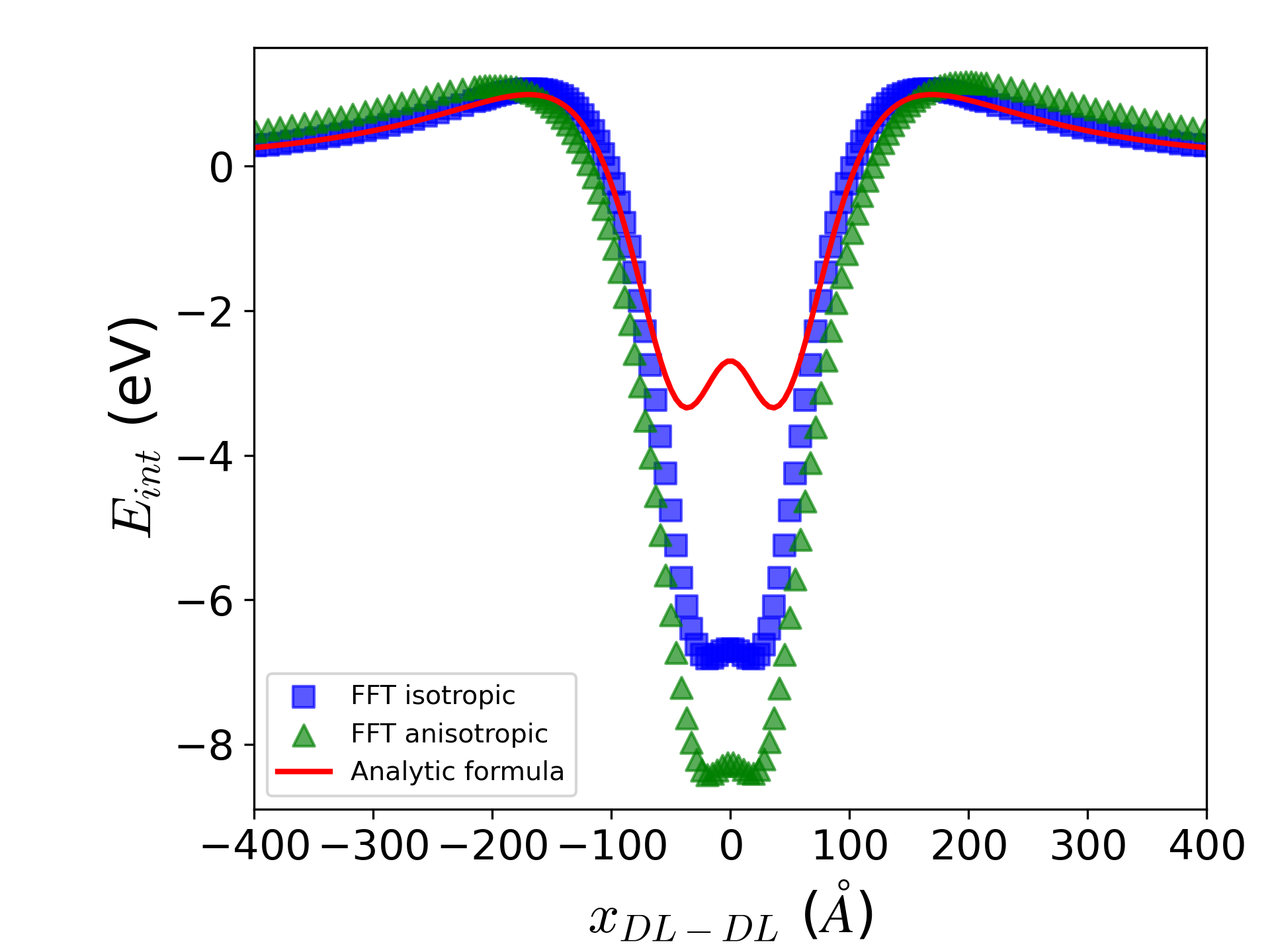}
 \caption{Left: Configuration of the two dislocation loops in the simulation box. Right: Elastic interaction energy of the two DL in iron as a function of their relative distance along the $x$ axis using an analytical expression from Ref. \cite{Dudarev2017} (red line), or using a full numerical evaluation of the interaction energy based on the FFT approach for both isotropic (blue squares) and anisotropic (green triangles) matrix.} 
 \label{fig:Eint_DL-DL}
\end{figure}

In order to calculate the elastic interaction energy between the two loops, one of the loops was fixed at the center of the box and the other one was placed at a distance $z_{DL_1} = z_{DL_2}$ + 100 {\AA} on the $z$ axis and its position on the $x$ axis was varied.  A general closed form expression for the elastic interaction of two small DLs based on the dipole approach was proposed by Dudarev and Sutton (Eq. (12) of \cite{Dudarev2017}). This expression is used here as the first method for computing the elastic interaction of the loops. The other two methods are the numerical evaluation of the interaction energy using the complete elastic fields created by the loops and computed using the FFT method for an isotropic matrix and for an anisotropic one. The resulting interaction energies are represented in Fig. \ref{fig:Eint_DL-DL}

As we can see, independently of the method used to calculate the interaction energy between the two loops, two minima are found. This result is in agreement with that obtained by McElfresh \textit{et al.} using a dislocation dynamic approach to calculate the hydrostatic stress between two loops in Mo for various loop separations\cite{McElfresh2022_2}.

\section{Object kMC model for the migration of defects under strain}
\label{sec:kMC_migration_strain}

In this section present the OkMC algorithm to simulate the kinetic of defects subjected to a strain field. The basis of the work starts from the parallel OkMC algorithm developed by Ortiz \textit{et al.}  \cite{Jimenez2016}. In contrast to the classical OkMC models \cite{BKL75,Gillespie76} (commonly referred as BKL) where only one particle can perform a single event during a time step, in this approach it is assumed that all moving defects are potentially able to undergo events during each time step $\delta t$. Thereby, the time step is fixed and its value is large enough so that a significant number of events can occur. This feature is fundamental for the coupling with FFT for interaction energy calculations, because it minimizes the number of times that elastic fields have to be recomputed/translated. Nevertheless,  $\delta t$ value must also be small enough so that the system as a whole does not change significantly and so that physical accuracy is preserved.

In order to comply with these two constraints, the time step $\delta t$ is taken as the inverse of the maximum event frequency $\nu_{max}$ in the system, 
\begin{equation}
 \label{eq:time_step}
 \delta t = \frac{1}{\nu_{max}}.
\end{equation}
This time step is independent of the number of particles in the system, in contrast to the BKL algorithm. This way, the system can evolve by time steps orders of magnitude larger \cite{Jimenez2016} than in classical OkMC models, strongly reducing the computational cost of kMC simulations.  Once the time step is fixed, the number of events $N_i$ that each defect $d_i$ of the system can undergo during $\delta t$ is determined using the Poisson distribution:

 \begin{equation}
  \label{eq:poisson}
  P(N_i, \nu_i; \delta t) = \frac{1}{N_i!}\left(\nu_i \delta t\right)^{N_i} \exp(-\nu_i \delta t)
 \end{equation}
where $\nu_i$ is the frequency of the event the defect $d_i$ can undergo.

\begin{figure}[ht]
\centering
\includegraphics[width=0.48\textwidth]{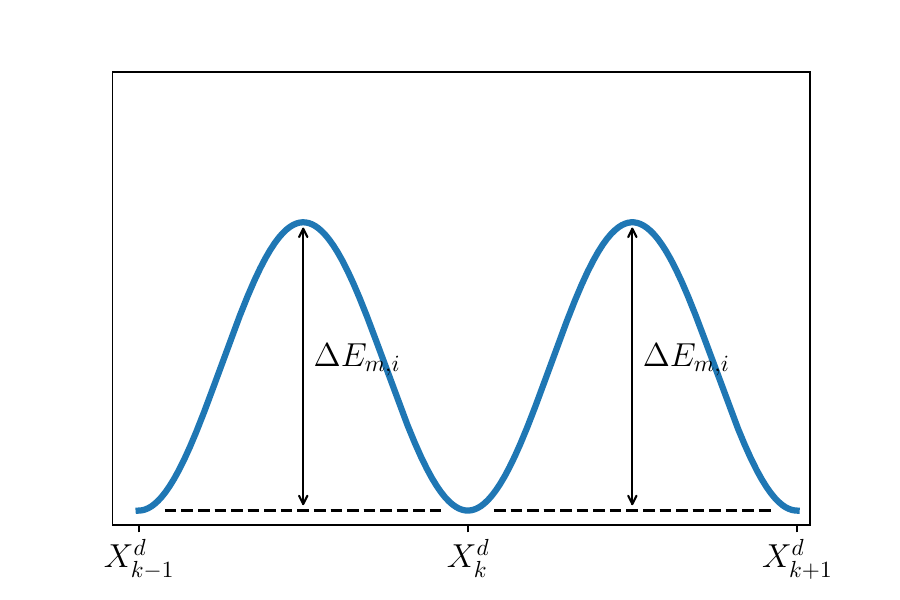}
\includegraphics[width=0.48\textwidth]{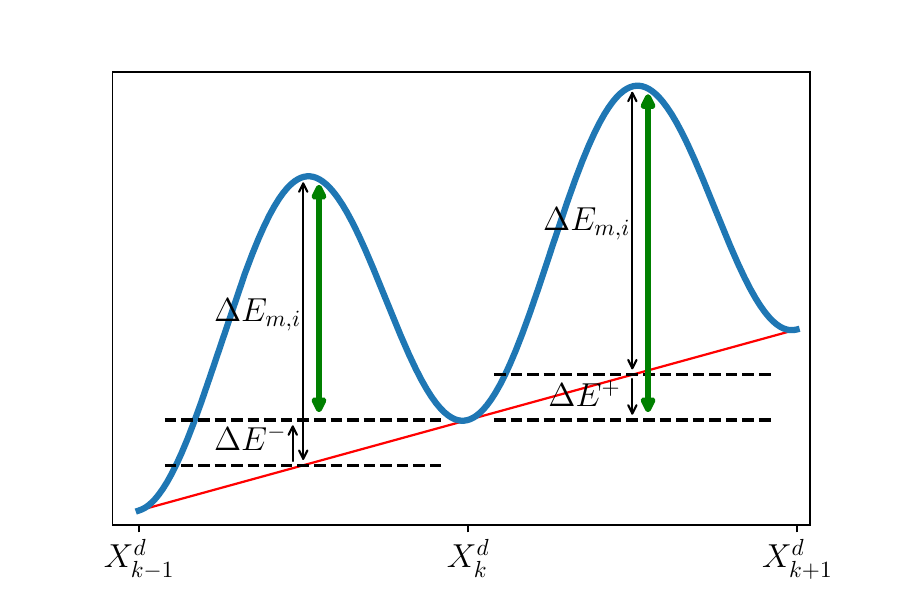}
 \caption{Energy barriers of a moving defect (a) without interaction energy (b) with interaction elastic energy increasing linearly to the right direction. In red the elastic energy, in blue the total energy and the green arrow indicates the effective barriers. In this case, the migration of the particle to position $X_{k-1}$ is more probable. }
 \label{fig:barriers}
\end{figure}

Due to the properties of the Poisson distribution and to the choice of the time step (Eq. \eqref{eq:time_step}), the fastest defects in the system --with frequency $\nu_{max}$-- perform in average one event per time step. In the scope of this research it is assumed that moving defects can perform only one type of event, a thermally activated migration process. The jump frequency for the defect $d_i$ is thus given by
\begin{equation}
 \label{eq:unbiased_boltzmann}
  \nu_i = \nu_0 \cdot \exp\left(-\frac{E_{m,i}}{k_BT}\right),
\end{equation}
where $\nu_0$ is the attempt frequency, $E_{m,i}$ the migration barrier and  $k_B$ is the Boltzmann constant. The exponential term represents the probability that the defect $d_i$ has a kinetic energy higher than $E_{m,i}$ and thus, can perform a jump. The migration energy $E_{m,i}$ is the maximum difference of potential energy that the defect must surmount when it jumps from one stable position at $X^d_k$ to another one (e.g, $X^d_{k+1}$) and it is assumed to be located halfway between the initial state at $X^d_k$ and the final state at $X^d_{k+1}$. In the absence of external elastic fields the migration barrier is the same in the different jump directions and therefore hops occur with equal probability in all directions. However, this no longer holds when defects are subjected to a strain field.  When a random walker evolves in a crystal which is elastically deformed by external loads or by the presence of other near defects, the energy barrier in Eq. \eqref{eq:unbiased_boltzmann} varies due to the elastic energy interaction. Let  $E(X^d_k)$ be the elastic energy of the system for the defect in its current position. When the defect attempts to jump from $X^d_k$ to a new position, $X^d_{k+1}$, besides the nominal migration energy $E_{m,i}$, it must surmount an additional difference of potential energy, $\Delta E^+$, defined halfway between $X^d_k$ and $X^d_{k+1}$ given by  $\Delta E^+=\frac{1}{2}(E(X^d_{k+1})-E(X^d_k))$. The frequency of forward jumps becomes thus:

\begin{equation}
 \label{eq:biased_forward_boltzmann}
  \nu_i^+ = \frac{1}{2} \cdot \nu_0 \cdot \exp\left(-\frac{E_{m,i} + \Delta E^+}{k_BT}\right),
\end{equation}
where the factor 1/2 comes from the fact that we only consider here forward hops. Similarly, we can define a backward jump frequency $\nu_i^-$ governing the rate of jumps in opposite direction, from $X^d_k$ to $X^d_{k-1}$
\begin{equation}
 \label{eq:biased_backward_boltzmann}
  \nu_i^- = \frac{1}{2} \cdot \nu_0 \cdot \exp\left(-\frac{E_{m,i} + \Delta E^-}{k_BT}\right)
\end{equation}
where $\Delta E^-$ is defined halfway between $X^d_{k-1}$ and $X^d_k$ and is given by $\Delta E^- = \frac{1}{2}(E(X^d_{k-1})-E(X^d_k))$. A sketch showing the modification of the energy landscape of a moving defect by the presence of an elastic interaction energy is represented on Fig. \ref{fig:barriers}.
If $\Delta E_{int}^+ > 0$ the effective migration energy to overcome by the defect becomes higher and, consequently, the jump becomes less probable than in the absence of the elastic energy, and the contrary effect happens if  $\Delta E_{int}^+ < 0$.  Same reasoning applies in the backward direction (from $X^d_k$ to $X^d_{k-1}$). 
As we can see, the main effect of the spatially-dependent elastic interaction energy is to affect the migration barrier of defects so the jump probabilities are biased as $\nu_i^- \neq \nu_i^+$, which can generate a drift of defects towards a preferential direction.

\subsection*{The OkMC algorithm}
\label{subsec:kMC_algorithm}

The algorithm of the OkMC including the FFT approach to compute elastic energy interactions is described here and summarized in Appendix \ref{AppendixA}.

First, the ensemble of defects and the frequency of the events they can perform is defined in order to determine the highest event frequency in the system and thereby, the time step with which the system will evolve during the simulation (see Eq. \eqref{eq:time_step}). For the sake of clarity and to focus on the effect of elastic interactions, it is assumed that defects do not change in size and/or type during the simulation and the possible reactions that could occur between two near defects are also not considered. Hence, the highest event frequency and the time step can be fixed once for all at the beginning of the simulation.
Here, it is important to note that in principle, the time step with which the system evolves should be updated as the global strain field changes with the position of the defects. Indeed, owing to Eqs. \ref{eq:biased_forward_boltzmann}-\ref{eq:biased_backward_boltzmann}, the frequency of events locally depends on the elastic energy landscape and so does the highest event frequency in the system. However, we found out that the $\Delta E_{int}^{+/-}$ are relatively small compared to the nominal migration energy $E_{m,i}$ of defects. Hence, the adaptive time step that would result is always close to the one calculated with the nominal migration energy. Therefore, for the sake of simplicity, we chose to use a constant time step throughout the simulations, the one calculated at $t=0$ and with the nominal migration energy.

Then, the elastic fields of the different type of defects that will evolve in the system are calculated following section \ref{sec:FFT_energies_2}. These fields $\boldsymbol{\EPS}^{d0,e}(\mathbf{x})$ are computed in a reference position and then stored in its Fourier representation. The elastic strain $\boldsymbol{\EPS}^e(\mathbf{x})$ of the whole system is then obtained by superposition, shifting the elastic strain field of each defect in the Fourier space (as explained in Sec. \ref{sec:FFT_energies_2}) to its coordinates $\mathbf{X}^d_i(t)$.

At that point, the evolution of the ensemble of defects interacting through their elastic fields starts. At the beginning of a time step $t$, the elastic energy of the system is known from the elastic strain $\boldsymbol{\EPS}^e(\mathbf{x})$ calculated with the current coordinates of the defects  $\mathbf{X}^d_i(t)$. Then, for each moving defect $d_i$, the number of hops it can perform during $\delta t$ owing to $\nu_i$ is calculated using the Poisson distribution defined by Eq. \eqref{eq:poisson}. 

If the defect $d_i$ must perform at least one jump during $\delta t$, then we must evaluate the biased frequencies corresponding to the different possible jump directions the defect can undergo. This implies that the variation of elastic potential energy $\Delta E_{int}$ (see Eqs. \eqref{eq:biased_forward_boltzmann} and \eqref{eq:biased_backward_boltzmann}) corresponding to each possible jump must be calculated. After all the biased jump frequencies $\nu_i^j$ have been calculated for the different possible jumps $j$ that the defect $d_i$ can undergo, the jump to be performed must be selected. For this, a cumulative function or frequency line is calculated as follows:

\begin{equation}
\label{eq:frequency_line}
 R_k^i = \sum_{j=1}^{k} \nu_i^j
\end{equation}
for $k=1,...,N_{\lambda}$, where $N_{\lambda}$ is the number of different possible directions the defect $d_i$ can jump to. By definition the maximum value of this cumulative function is $R_{tot}^i=\sum_{j=1}^{N_{\lambda}} \nu_i^j$. 

A random number $0 < \xi \leq 1$ is chosen and the jump $j$ to perform is selected such that:
\begin{equation}
\label{eq:selection_jump}
\frac{R_{j-1}}{R_{tot}} < \xi \leq \frac{R_j}{R_{tot}}
\end{equation}

Once all the defects have been parsed, the elastic strain $\boldsymbol{\EPS}^e(\mathbf{x})$ of the system is updated taking into account the new coordinates of all the defects. Finally, the clock time of the system is incremented by $\delta t$. This loop is repeated until the final time has been reached.

\section{Results}
\label{sec:simul_cases}

In this section we shall explore different cases of evolution of defects in Fe interacting through elastic forces. Both self-interstitials and prismatic dislocation loops (DL) will be considered. Experimental observations show that in body-centered cubic Fe, DLs can form with two different Burgers vectors\cite{Yao08, Prokhodtseva13}, $1/2\langle 111 \rangle$ or $\langle100\rangle$. While $1/2\langle 111 \rangle$ loops migrate almost athermally by 1D glide along the direction of their Burgers vector with a  small migration barrier of $\sim$ 0.1 eV \cite{Terentyev07,Willaime05,Jansson13}, electron microscopy observations show that $\langle100\rangle$ loops only start migrating above 770 K\cite{Arakawa06}. Therefore, they are often considered as immobile. For these reasons, in what follows, we only consider the evolution of $1/2\langle 111 \rangle$ DLs. In the rest of the section, DLs will simply refer only to $1/2\langle 111 \rangle$ DLs. The DLs will have the same characteristics as in subsections \ref{subsec:Eint_DL_disloc} and \ref{subsec:Eint_DL_DL}. 

Simulations are performed in a cubic box of side 500 $a_0$ with periodic boundary conditions, $a_0$ being the lattice parameter of bcc Fe with an accepted value of approximately 2.83 {\AA} (see Refs. \cite{Ma2019,Haas2009}). The box was defined with a size large enough so as to avoid a DL could interact with itself through the boundary conditions. The coordinate system is defined such as the Burgers vector of the DL ($1/2\langle 111 \rangle$) is parallel to the $x$ axis. The two other axes ($y$ and $z$) are defined in order to obtain an orthogonal reference system. For the migration energy of the DLs in Fe, a value of 0.1 eV is taken, a commonly accepted value in the community\cite{Malerba21,Jansson13}. A temperature of 300 K (room temperature) was used in the simulations. At this temperature, the characteristic time step for the jump of a DL, given its migration energy in Fe, is estimated to be about 4.78 ps. The evolution of the DLs was simulated for 500 steps in all simulations reported below. All the results shown below were obtained with the FFT method taking into account the anisotropy of Fe. Since the kMC method is stochastic by nature, 10 runs were performed in each condition in order to obtain a statistical trend of the DL trajectories.

\subsection{Two dislocation loops migrating and interacting}
\label{subsec:evol_2_DLs}

The dynamics of two DLs migrating and interacting in Fe is simulated for different initial relative positions. The general configuration of the DLs is represented in the Fig. \ref{fig:Eint_DL-DL} (Left), where their Burgers vector is also reported (arrows).The evolution of their position along their axis of motion ($x$) is monitored during the simulation. The analysis of the interaction energy of two DL was done in section 4.4 and the results obtained using anisotropic matrix in FFT were represented in Fig. \ref{fig:Eint_DL-DL} (Right, green triangles) as a function of their relative distance $x_{relat} = x_{DL_1} - x_{DL_2}$ for a horizontal separation $z_{relat}$ = 100 {\AA}.

\begin{figure}[ht]
\centering
\includegraphics[scale=0.40]{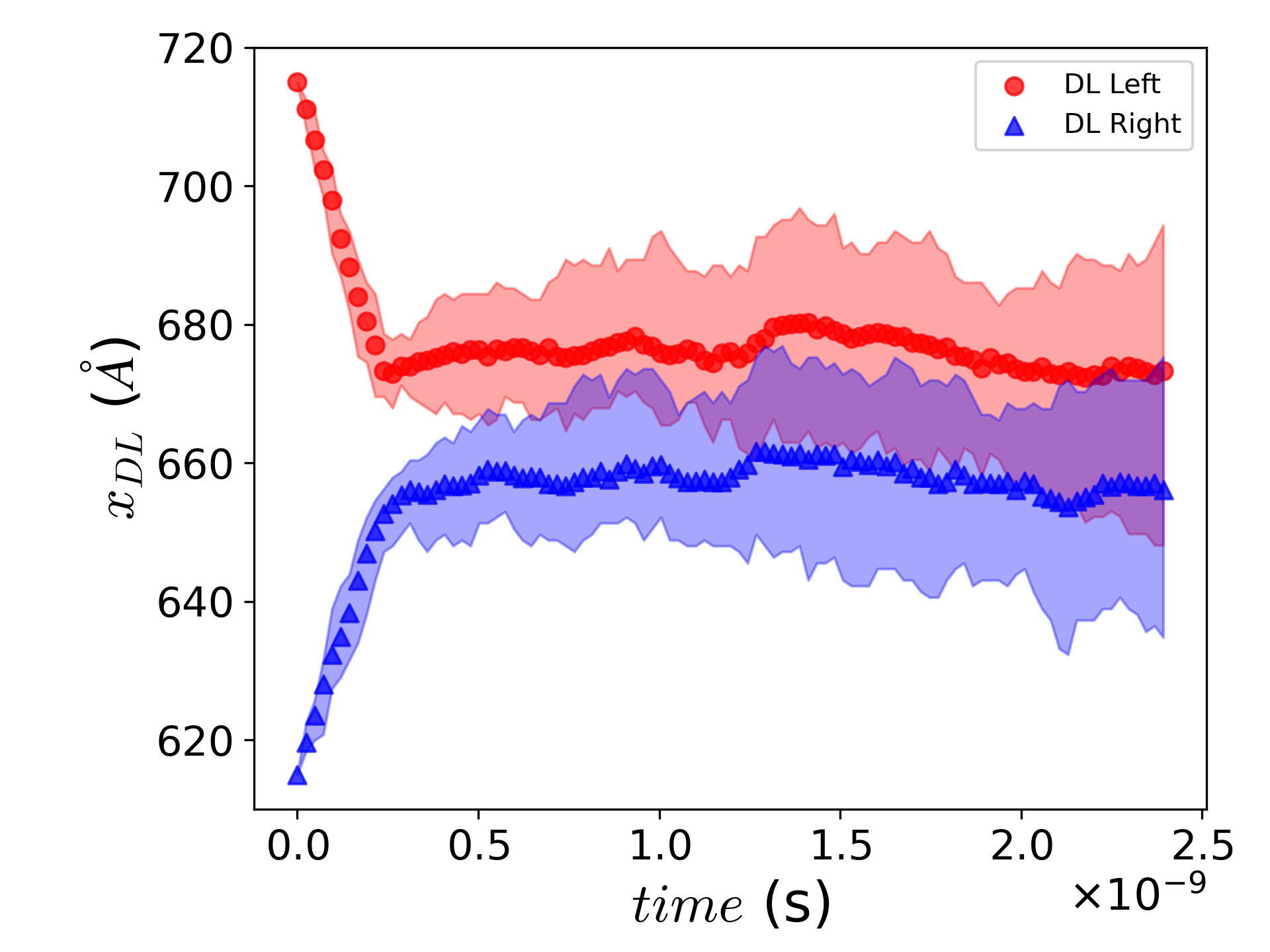}
\includegraphics[scale=0.40]{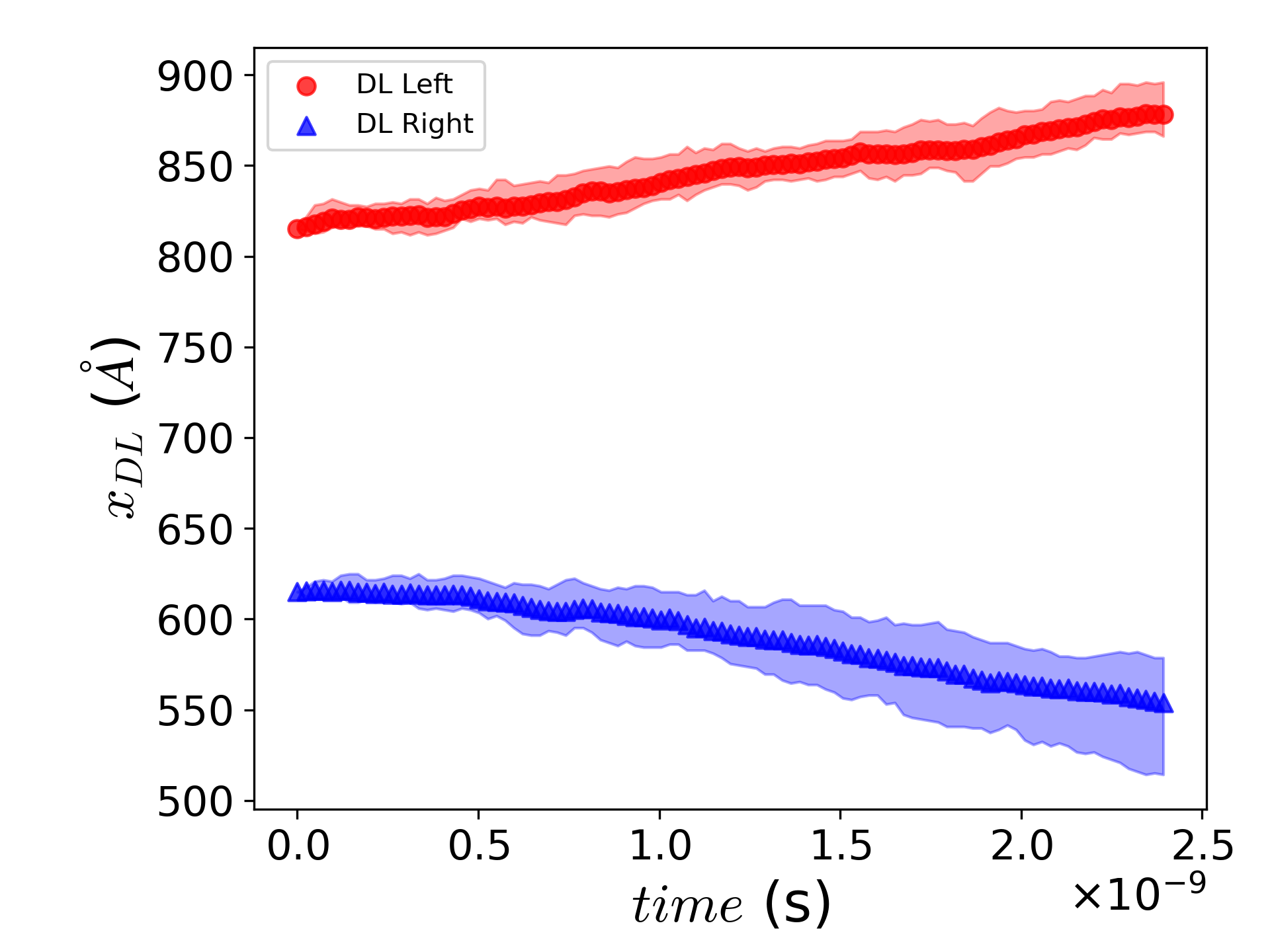}
 \caption{Left: Evolution of the coordinates of the loops along the $x$ axis as function of time in the case where they are in the potential well ($x_{relat} = 100$ {\AA}) seen in Fig. \ref{fig:Eint_DL-DL}. Right: Evolution of the coordinates of the loops along the $x$ axis as function of time in the case where they are outside the potential well ($x_{relat} = 200$ {\AA}) seen in Fig. \ref{fig:Eint_DL-DL}. Symbols represent the mean trajectories obtained over the 10 runs. The envelope of the trajectories of all the loops is illustrated by the corresponding colored areas.}
 \label{fig:trajectory_2DLs}
\end{figure}

Two cases are considered. First, the dynamics of two loops at an initial distance $x_{relat} = 100$ {\AA} is studied such that they are in the potential well though not yet at the minimum of the elastic energy. The resulting evolution of the mean trajectories (along the $x$ axis) of the loops are reported as a function of time in Fig. \ref{fig:trajectory_2DLs} (Left). In the second case, the loops are initially separated by a relative distance $x_{relat} = 200$ {\AA}, i.e., outside the region of the potential well according to Fig. \ref{fig:Eint_DL-DL}. The evolution of the mean trajectory of the loops corresponding to this initial separation is represented in Fig. \ref{fig:trajectory_2DLs} (Right). In both cases, the envelope of all the possible trajectories obtained over the 10 runs is illustrated in order to evidence their trend.

\subsection{Two dislocation loops migrating and interacting with a dislocation line}
\label{subsec:evol_2_DLs_1DL}
In this subsection, the OkMC-FFT code is used to predict the dynamics of two dislocation loops interacting near a dislocation line in Fe. This is a case of interest as it is experimentally observed that DLs can pin to dislocation lines, which is believed to be responsible for the hardening of steels under irradiation \cite{Hardie13, Dethloff18}. For this, in the system of coordinates described at the beginning of the section, we defined an edge dislocation of type $\frac{1}{2}(0\bar1 0)[111]$, i.e., with the same characteristics as in Sec. \ref{subsec:Eint_SIA_disloc}. Considering the absence of external loading, and with the aim of studying the dynamics of the loops around the dislocation isolated from other phenomena, it is assumed that the dislocation line is immobile during the whole simulation time.

\begin{figure}[ht]
\centering
\includegraphics[scale=0.5]{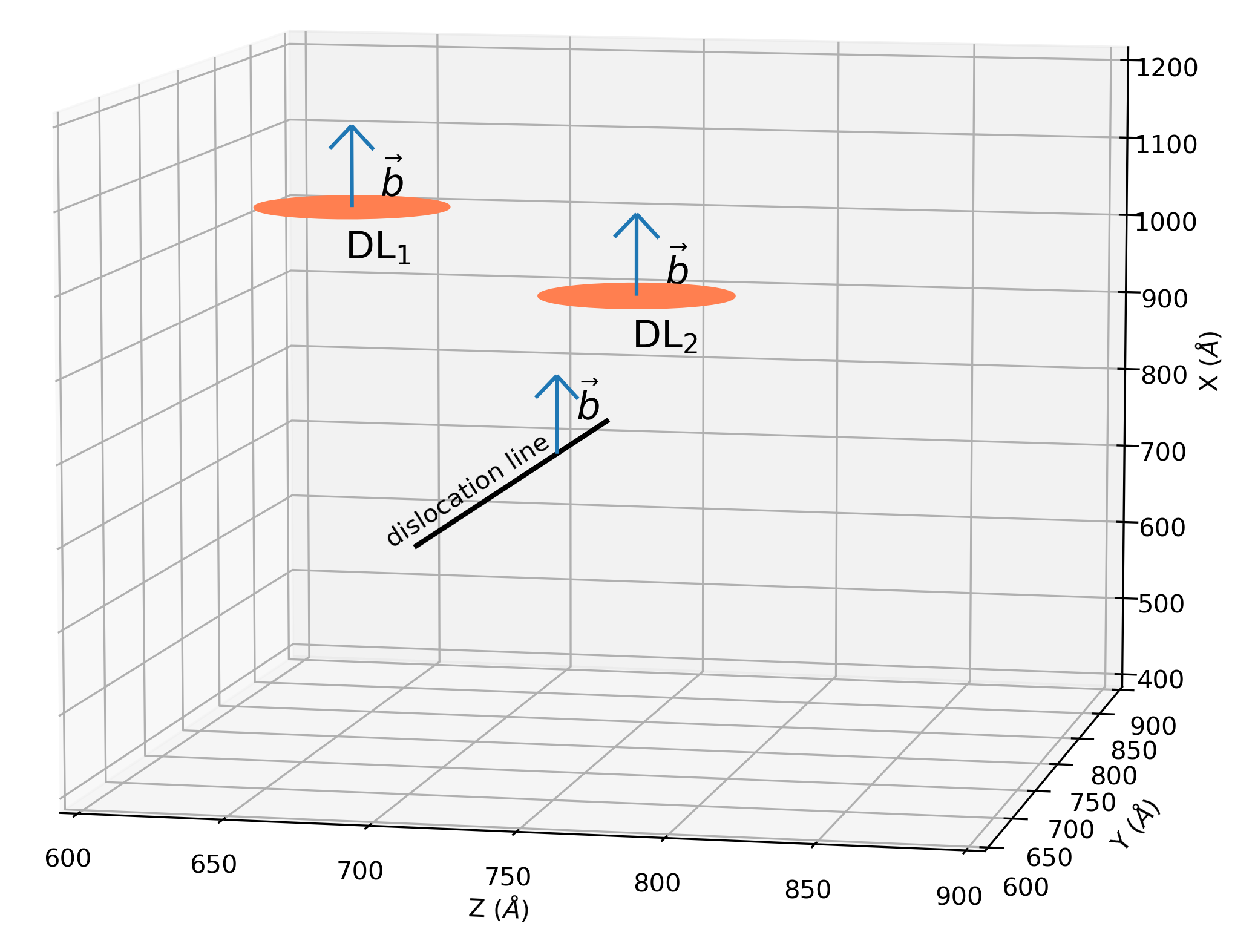}
 \caption{Configuration of the straight dislocation line and two prismatic loops in the simulation box.} 
 \label{fig:3D_2_DLs_disloc}
\end{figure}

As in the first case studied in the previous subsection, we consider the case of two DLs separated by a horizontal distance $z_{relat} = 100$ {\AA} and by a relative distance $x_{relat} = 100$ {\AA} from each other. The distance between the two loops is thus so that they are in the potential well (Fig. \ref{fig:Eint_DL-DL}), i.e., in the attraction zone. Here, the loops were located symmetrically on the $z$ axis on both sides of the dislocation line at a distance of 50 {\AA} each, as shown in Fig. \ref{fig:3D_2_DLs_disloc}.

\begin{figure}[ht]
\centering
\includegraphics[scale=0.40]{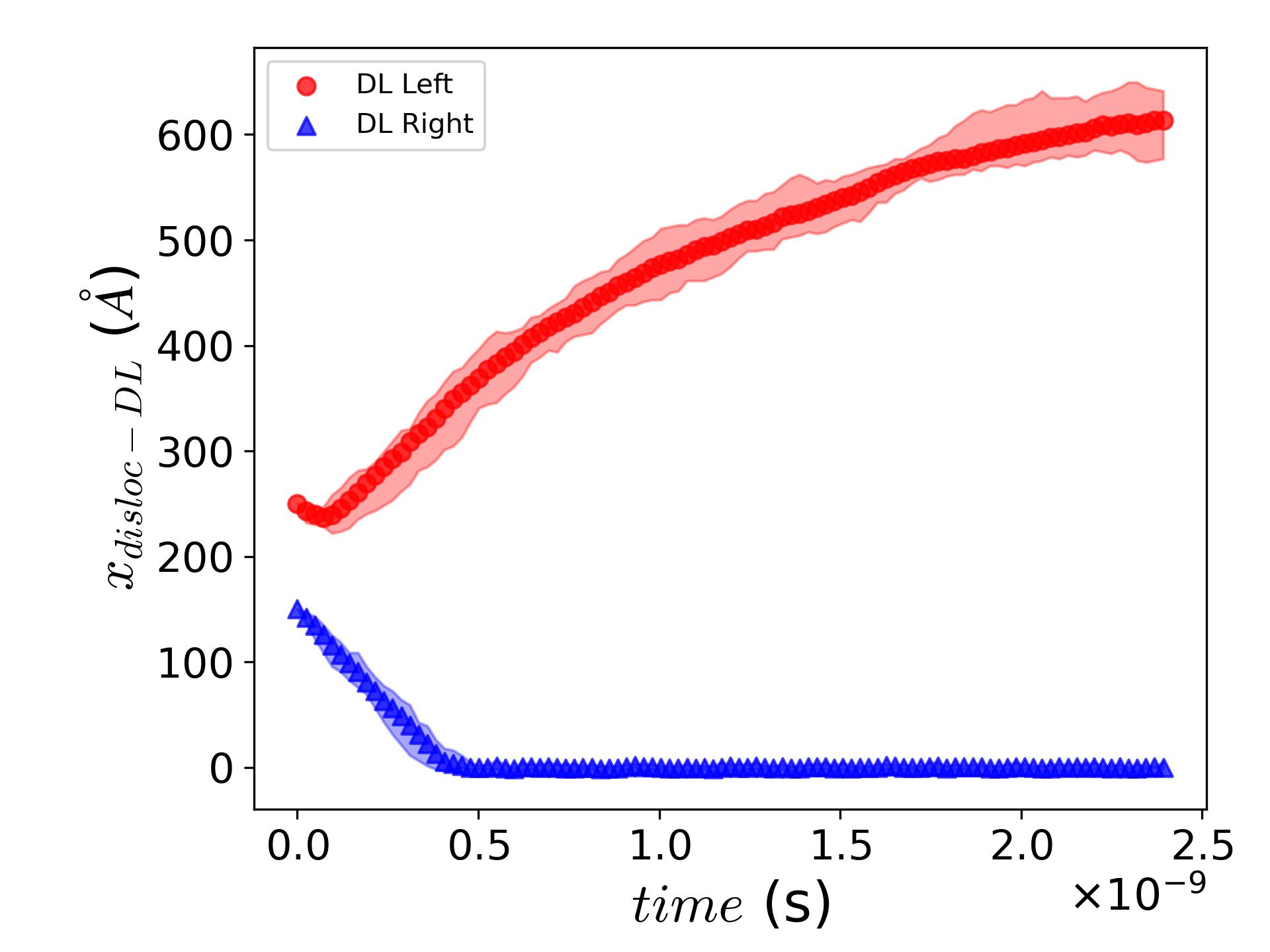}
\includegraphics[scale=0.40]{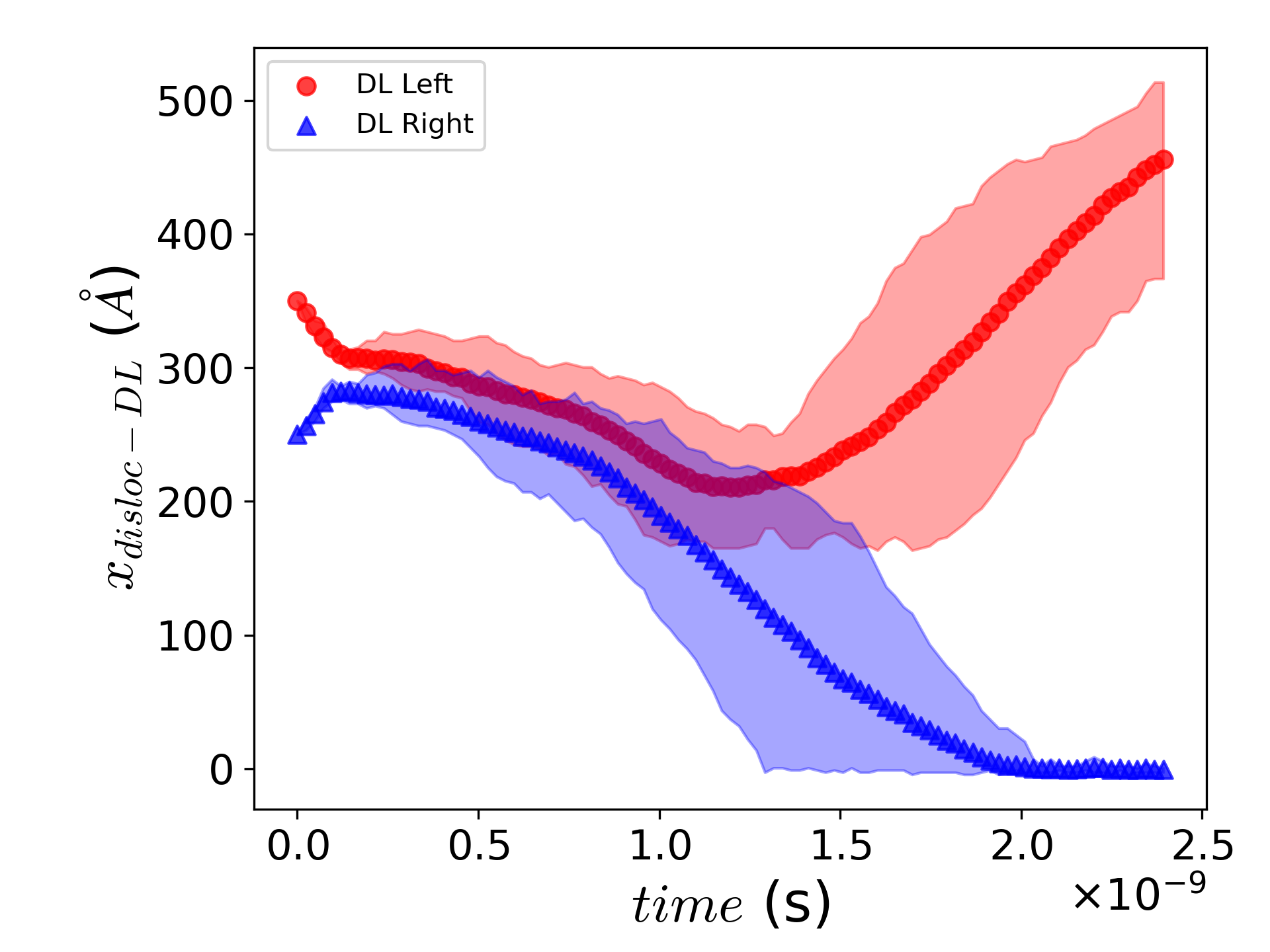}
 \caption{Left: Evolution of the coordinates of the loops along the $x$ axis as function of time in the case where their center of mass is close to the dislocation line (200 {\AA}). Right: Evolution of the coordinates of the loops along the $x$ axis as function of time in the case where their center of mass is far from the dislocation line (300 {\AA}).  Symbols represent the mean trajectories obtained over the 10 runs. The envelope of the trajectories of all the loops is illustrated by the corresponding colored areas.}
 \label{fig:trajectory_2DLs_disloc}
\end{figure}

First,  the evolution of the loops is simulated when their center of mass on the $x$ axis is at a relatively short distance from the dislocation line, specifically at 200 {\AA}. The relative distance of each loop to the dislocation along the $x$ axis is represented as function of time in Fig. \ref{fig:trajectory_2DLs_disloc} (Left). The second case corresponds to  two loops with center of mass is relatively far from the dislocation line, 300 {\AA}. The trajectories of the loops along the $x$ axis obtained in this case with the OkMC-FFT can be seen in Fig. \ref{fig:trajectory_2DLs_disloc} (Right). In both figures the envelope of all the trajectories is included.

\subsection{SIAs migrating in the presence of a Dislocation Loop trapped near a dislocation line}
\label{subsec:SIAs}
\begin{figure}[ht]
\centering
\includegraphics[scale=0.30]{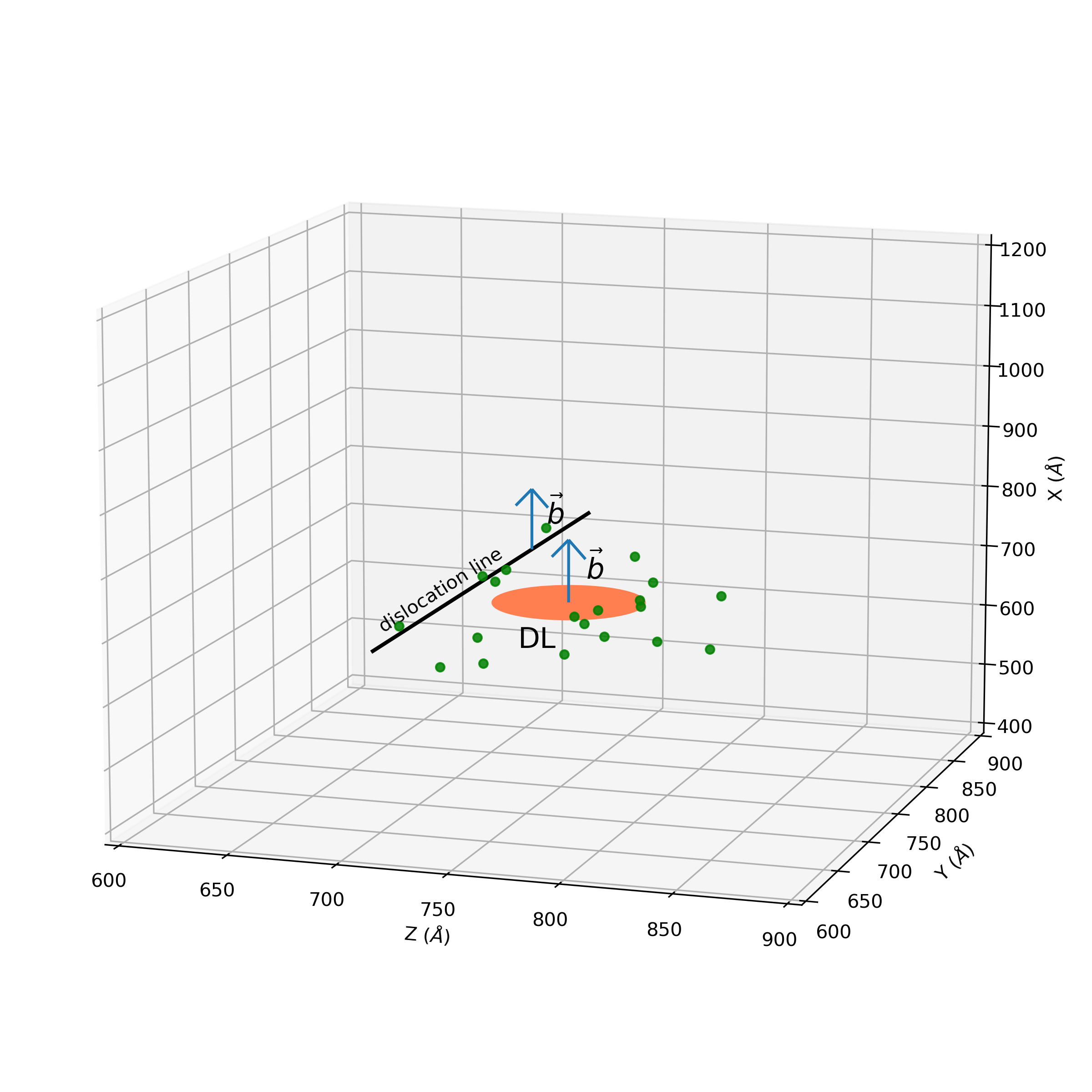}
\includegraphics[scale=0.30]{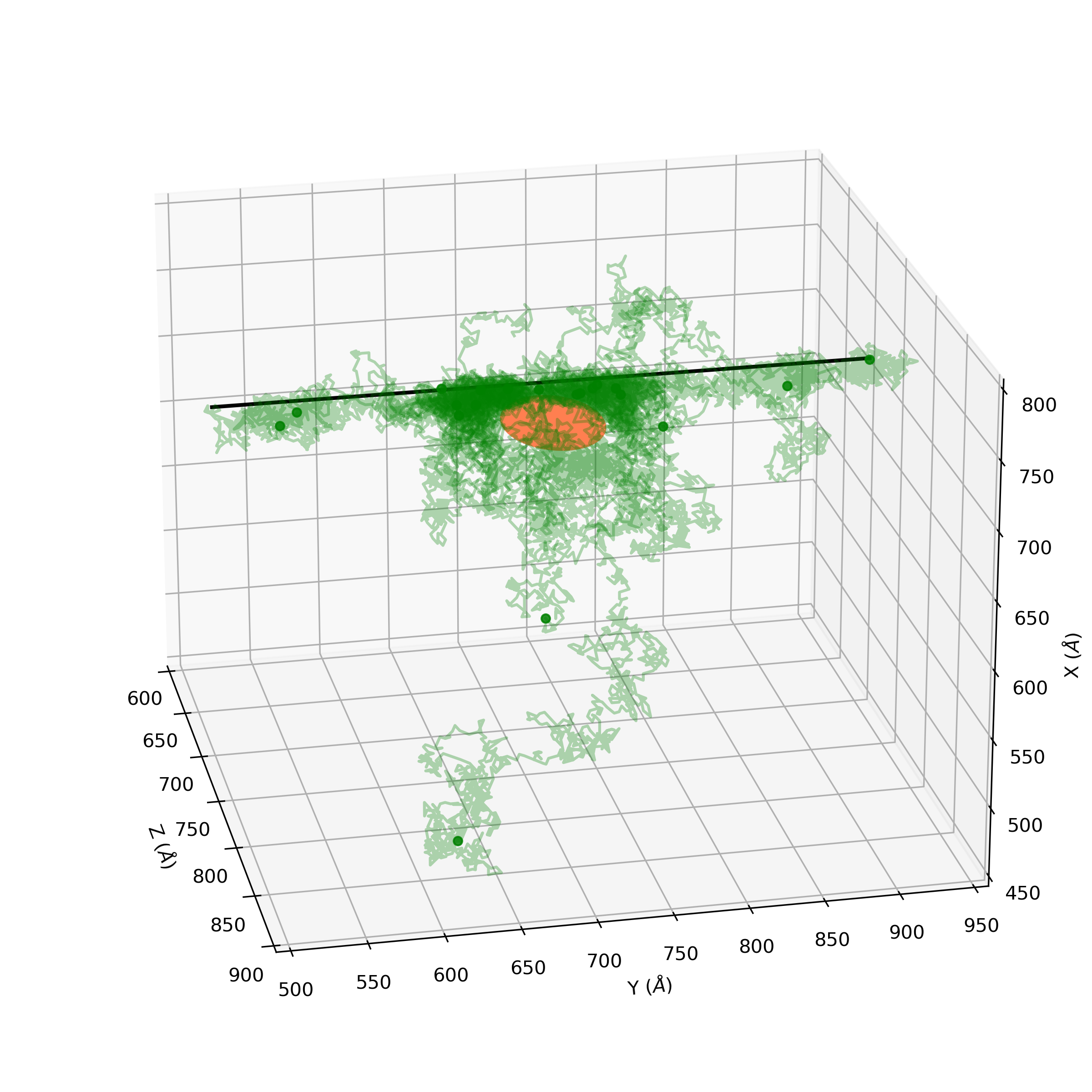}
 \caption{Left: Initial distribution of SIAs in the presence of a dislocation loop trapped near a dislocation line in the simulation box. Right: Trajectories and final positions of SIAs after 2000 steps.}
 \label{fig:SIAs_trapped_DL}
\end{figure}

In the last subsection, simulations showed that under certain conditions a DL can be trapped and become immobile near an edge dislocation line. This occurs because of the deep potential well of elastic energy that establishes between the dislocation line and the loop and from which this latter cannot escape (see Fig. \ref{fig:Eint_disloc-DL} (green triangles)). When this occurs, it becomes interesting to study how point defects such as SIAs evolve near the trapped loop. In order to investigate the interaction of mobile SIAs with the strain field of a $1/2\langle 111 \rangle$ loop trapped near a dislocation line in Fe, an edge dislocation and a DL with the same characteristics as in previous subsection are defined. The DL is placed at a position where in principle it should remain trapped owing to previous simulation. To achieve this, the DL was located in front of the dislocation line ($x_{disloc-DL} = 0$) and at a distance $z_{disloc-DL} = 50$ {\AA}, i.e., exactly at the bottom of the potential well shown in Fig. \ref{fig:Eint_disloc-DL} (green triangles). As a way to study the evolution and interaction of SIAs with an immobile DL, we defined a cubic box with sides of 150 {\AA} centered on the coordinates of the DL and containing 20 SIAs randomly distributed. This corresponds to a concentration of approximately $6\cdot 10^{18}$ cm$^{-3}$, a realistic value under irradiation conditions. These initial conditions are represented in Fig. \ref{fig:SIAs_trapped_DL} (Left).
For the migration energy of the SIAs in Fe,  the value commonly accepted in the literature \cite{Malerba21, FuNature2005} of 0.34 eV was chosen. In these simulations, since SIAs are point-defects with very localized strain fields, the dipole tensor approximation is perfectly justified and is used to calculate their interaction energy in an external elastic field using Eq. \eqref{eq:interaction_energy4}. The dipole tensor used is the one corresponding to the stable $<110>$ dumbbell SIA in Fe obtained by Ma and Dudarev using density functional theory \cite{Ma2019} and reported in Table \ref{tab:Fe}. Since the components of the dipole tensor of the $<110>$ dumbbell in Fe change along the migration pathway (see Fig. 7 of Ref. \cite{Ma2019}), the dipole tensor at the saddle point (taken from Ref. \cite{Ma2019}) was taken into account to correctly calculate the interaction energy at halfway. Here, the agglomeration of SIAs to the DL was ignored in the simulations. The authors are aware that in principle, the SIAs should disappear as they agglomerate to the DL \cite{Ortiz07,Malerba21} but for the sake of clarity, this mechanism was not taken into account here. Our main goal here is to evidence how the SIAs evolve in the strain field induced by the trapped DL.

The evolution of the SIAs was simulated at room temperature and for 2000 kMC steps. Since SIAs are able to migrate in 3D, in Fig. \ref{fig:SIAs_trapped_DL} (Right) we report their trajectory around the DL in the ($xyz$) plot. The dislocation loop and the dislocation line were also illustrated to better evidence the trajectory of the SIAs and their preferential migration path.

\section{Discussion}

\subsection{Interaction energies}
The results presented in section 4 showing the interaction energies of different defects obtained using the FFT proposed approach and compared with the dipole approximation will be discussed here.

In the case of the interaction between a SIA and a straight dislocation (section \ref{subsec:Eint_SIA_disloc}), in Fig. \ref{fig:Eint_disloc-SIA} it can be observed that the interaction energy provided by the analytical expression based on a dipole and the Volterra strain field predicts a result superposed to the one obtained solving the dislocation field by FFT assuming isotropy. This fact further validates the FFT based field dislocation mechanics approach to compute the strain field generated by a straight dislocation line. Only very small differences are found at the edges of the cell due to the periodicity condition in FFT, in contrast to the dislocation in an infinite medium considered by Volterra. On the contrary, non-negligible differences are found when these curves are compared with the FFT results taking into account the anisotropy, in particular when the SIA is close to the dislocation line ($x \simeq 0$). This shows the limitations of using approaches based on the isotropic elastic response in a material with considerable high anisotropy such as iron. Moreover, it is remarkable that the computational cost of using an anisotropic medium is the same as for an isotropic one.

Regarding the interaction energy of a dislocation loop with a straight dislocation, Fig \ref{fig:Eint_disloc-DL} show that both methods predict the same interaction energy when the loop is far from the dislocation. This of course is expected since far from the dislocation, the elastic field that the loops feels varies slowly and a mean value can be used. However, the Fig. \ref{fig:Eint_disloc-DL} evidences that the results obtained with both methods start to significantly differ as the DL gets close to the dislocation. This is due to the fact that the dipole approximation is no longer valid for defects with a non-negligible volume that evolve in a rapidly changing strain field, which is the case of a loop near a dislocation line. The FFT approach, which relies on the \emph{exact} evaluation of the interaction energy by integration, provides here up-to double depth of the energy minimum than the one predicted by dipole approach which will have a strong impact in the stability of the DL position near a dislocation.

Finally, in the  case of two interacting dislocation loops (section \ref{subsec:Eint_DL_DL})  the analytical approach reproduces with a good accuracy the values predicted by the FFT method both for the isotropic and anisotropic matrix when the two loops are relatively far from each other (Fig. \ref{fig:Eint_DL-DL}) . However, the analytic expression significantly differs from the results obtained with the FFT method as the loops get close. This difference is even more pronounced when the anisotropic elasticity of Fe is taken into account, providing the \emph{exact} evaluation through the FFT energy minima up-to three times higher than the analytical approximation. 

These result shows again the limitations of analytical expressions based on the dipole approximation under certain conditions. Clearly, the dipole approximation fails when the elastic field changes rapidly in space, as expected, and the FFT method should be used instead.

\subsection{Evolution of irradiation defect ensembles in Iron}

\subsubsection*{Migration of two interacting dislocation loops}
The trajectories of two dislocation loops interacting elastically presented in section \ref{subsec:evol_2_DLs} will be analyzed. In the first case (Fig. \ref{fig:trajectory_2DLs}, left), the distance between the two loops is in the potential well (distance of 100{\AA}). In this case, in a first stage, the loops move towards each other. Then, in a second phase, when the loops are at a certain relative distance along the $x$ axis, their trajectories become correlated and they seem to migrate as a single entity. Since the jumps for which $\Delta E_{int} < 0$ (condition fulfilled for the initial distance, Fig. \ref{fig:Eint_DL-DL} (Right)) are more probable as the resulting effective migration energy is lower (see Eqs. \eqref{eq:biased_forward_boltzmann} and \eqref{eq:biased_backward_boltzmann}) it is very likely that at the beginning of their evolution, the loops perform jumps towards the bottom of the potential well. Thereby, during the first phase, the loops tend to migrate towards each other until they reach one of the local minima of the interaction energy, a relative distance of approximately 20 {\AA} according to Fig. \ref{fig:Eint_DL-DL} (Right), which is confirmed by the kMC results seen in Fig. \ref{fig:trajectory_2DLs} (Left). Then, when one of the loops performs a random jump equilibrium condition is lost and the jump probabilities become biased again. In order to minimize the energy of the system the other loop performs a jumps towards the first loop. As a result, the loops drag each other, migrating apparently as a single entity. This explains the correlated trajectories of the loops and why they remain at a more or less constant relative distance during their evolution. This result is in very good agreement with what was experimentally observed by Dudarev \textit{et al.}\cite{Dudarev2010} in irradiated Fe. The authors showed using \textit{in situ} electron microscopy that the elastic interaction between two $1/2\langle 111 \rangle$ loops in Fe can be so strong that their dynamics becomes correlated and they can migrate as a single entity.

In the second case considered, Fig. \ref{fig:trajectory_2DLs}, right, the distance between their initial location is 200{\AA}. The trajectories in this case reveal that the loops tend to move in opposite directions. The reason is that for this initial separation the interaction energy (Fig. \ref{fig:Eint_DL-DL}) decreases by increasing their relative distance and therefore the loops tend to jump towards larger relative distances in order to minimize the interaction energy. In other words, the loops repeal each other, showing that the coordinated movement occurs only for initially nearby loops.

\subsubsection*{Migration of two interacting dislocation loops near an immobile dislocation}
The simulations of the interaction between two loops and a straight dislocation presented in section \ref{subsec:evol_2_DLs_1DL} were performed for two initial conditions. The first case was done for two loops having their center of mass on the $x$ axis at 200 {\AA} from the dislocation line, Fig. \ref{fig:Eint_disloc-DL}(Left) . This figure evidences that the loops follow a distinct and uncorrelated evolution in these conditions. The loop located on the left of the dislocation (DL$_1$) moves away from it whereas the loop located on the right (DL$_2$) migrates towards the dislocation and after some time, seems to stay trapped in front of the dislocation ($x_{disloc-DL} = 0$) as its position barely changes with time. The interaction energy of the dislocation line and a DL was studied in section 4.3 and the resulting interaction energies for a loop located on the right of the dislocation line (in z direction) as function of the position in $x$ direction is represented in Fig. \ref{fig:Eint_disloc-DL}. The interaction energy of the loop on the left of the dislocation is simply the opposite. Since the loop located on the left of the dislocation line is in a repulsion zone (interaction energy with the dislocation line decreases with the distance), it will tend to migrate away from it (red circles in Fig. \ref{fig:trajectory_2DLs_disloc} (Left)). On the other hand, the loop that is on the right of the dislocation line is in an attraction zone and hence falls in the potential well until it reaches the minimum of elastic energy, in front of the dislocation line ($x_{disloc-DL} = 0$). This explains why its position barely changes once it has reached this position on the $x$ axis (blue triangles in Fig. \ref{fig:trajectory_2DLs_disloc} (Left)). This is in agreement with the decoration of dislocations by loops that is observed experimentally and that was also predicted theoretically \cite{Trinkaus1997, Wen2005}.From this study we can conclude that, even though the interaction between the two loops is such that they should attract each other, as it was shown in the first case studied in previous subsection (see Figs. \ref{fig:Eint_DL-DL} and \ref{fig:trajectory_2DLs} (Left)), when they are relatively close to an edge dislocation line, their dynamics are mainly governed by their individual interaction energy with the dislocation line. This is actually expected since the attraction between a dislocation loop and a dislocation line is much stronger than that of two dislocation loops, as can be seen by comparing Figs. \ref{fig:Eint_disloc-DL} (Right) and \ref{fig:Eint_DL-DL} (Right).

The second case (Fig. \ref{fig:trajectory_2DLs} (Left)) corresponds to the two loops having their center of mass on the $x$ axis at 300 {\AA} from the dislocation line, where less influence is expected. Here, the resulting dynamics of the loops is more complex than in the previous case and this time exhibits distinct stages. First (until $t=1$ns), the loops follow a similar evolution as that shown in Fig. \ref{fig:trajectory_2DLs} (Left), that is, when two isolated loops move towards each other due to their mutual interaction energy. It is worth noting that the loop located on the left of the dislocation, i.e., that should be repealed by the dislocation (see Fig. \ref{fig:trajectory_2DLs_disloc} (Left)), this time migrates towards the dislocation. Similarly, we can see in Fig. \ref{fig:trajectory_2DLs_disloc} (Right) that the loop on the right and that should be attracted by the dislocation according to Figs. \ref{fig:Eint_disloc-DL} (Right) and \ref{fig:trajectory_2DLs_disloc} (Left), tends to move away from it and migrates towards the other loop. This suggests that in these conditions, i.e., when the two loops are relatively far from the dislocation line, their mutual interaction prevails over the interaction DL--dislocation and governs their dynamics. Then, similarly to the first case studied in previous subsection, when the loops are at a certain distance from each other, which corresponds to one of the local minima predicted in Fig. \ref{fig:Eint_DL-DL} (Right), we observe that their migration becomes correlated and they move as a single entity by dragging each other (see previous subsection). Furthermore, we can see that the pair of loops does not wander as a pure random walker but clearly drifts towards the dislocation line. This suggests that the loop on the right is attracted by the dislocation line and drags the other one during its migration. After this phase of their dynamics, it  can be observed in Fig. \ref{fig:trajectory_2DLs_disloc} (Right) that the pair of loops breaks up and their evolution becomes independent. From this moment, the loops follow a similar evolution as that shown in Fig. \ref{fig:trajectory_2DLs_disloc} (Left) and corresponding to the evolution of two loops close to the dislocation line. The trajectory of the loop located on the left of the dislocation abruptly changes and the loop starts moving away from the dislocation, indicating a strong repulsion. At the same time, the loop located on the right continues its migration towards the dislocation line until its position no longer changes. This evidences that, as the pair of loops approaches the dislocation line, their evolution is no longer dominated by their mutual interaction but starts to be governed by their individual interaction with the dislocation line.

\subsubsection*{SIAs migrating in the presence of a Dislocation Loop trapped near a dislocation line}

The interaction of SIAs in the presence of a $1/2\langle 111 \rangle$ loop trapped near an edge dislocation in iron was presented in section \ref{subsec:SIAs} and the trajectories followed are represented in Fig. \ref{fig:SIAs_trapped_DL}. In this figure, it can be observed that the space above and below the DL is clearly depleted by the SIAs that seem to move away from this region. On the other hand, most of the SIAs tend to move towards the habit plane of the DL and in the space between the loop and the dislocation line, where they seem to concentrate. Only some SIAs have escaped from the volume where they were initially located.

\begin{figure}[ht]
\centering
\includegraphics[scale=0.7]{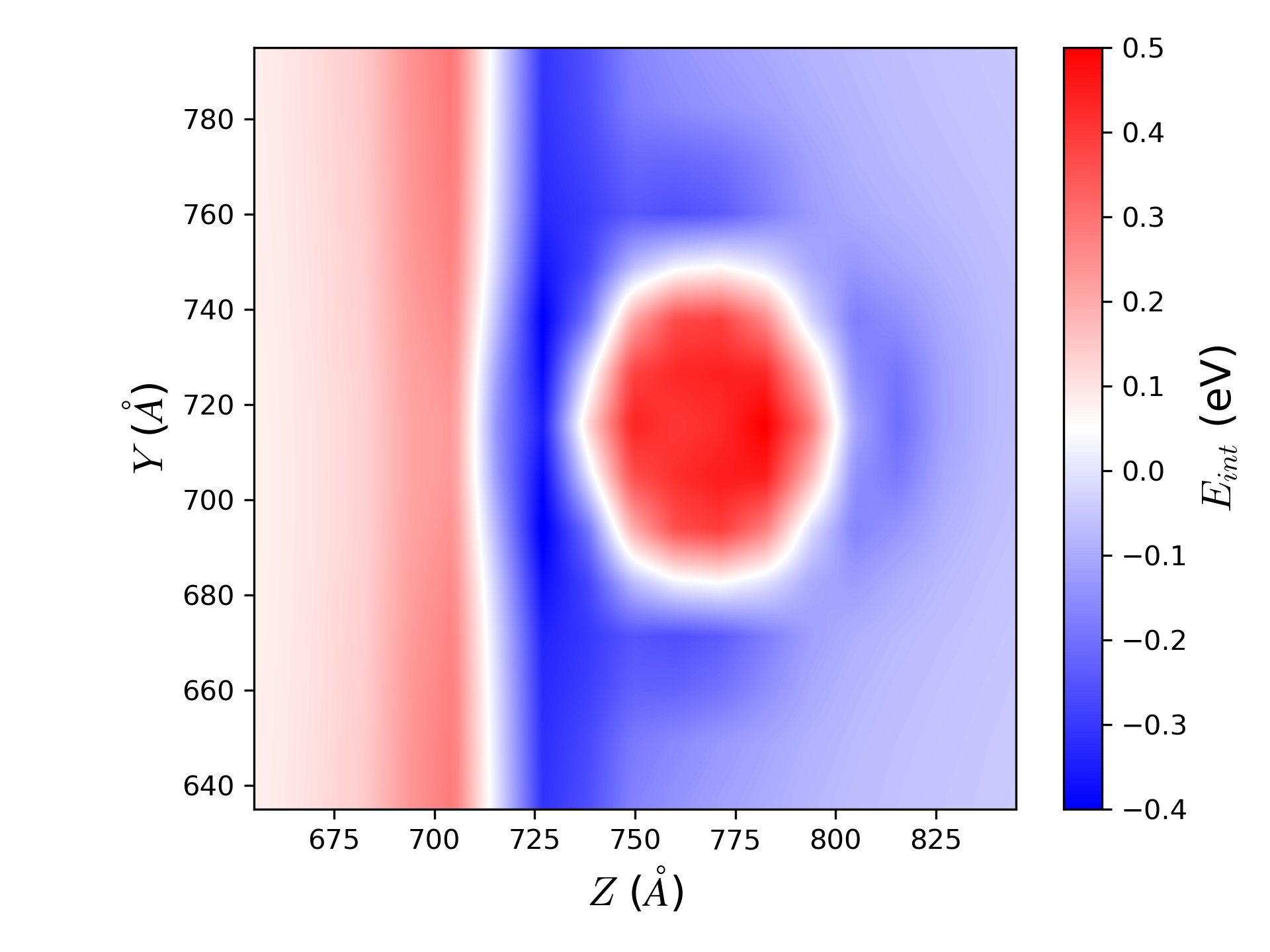}
 \caption{Map of the elastic interaction energy of a SIA in the $yz$ plane in the presence of a dislocation loop trapped near an edge dislocation line.} 
 \label{fig:Eint_2D_SIA_trapped_DL}
\end{figure}

In order to understand the mechanism responsible for the clearly non-random trajectory followed by the SIAs near the trapped DL, in Fig. \ref{fig:Eint_2D_SIA_trapped_DL} the interaction energy of a SIA induced by the trapped DL and the dislocation line is represented in the plane ($yz$) of the DL. In this Fig., we easily discern where the dislocation line is located, along the $y$ axis at $z=715$ {\AA}. The region on the left of the dislocation line is clearly a zone of repulsion for the SIAs as the interaction energy decreases as the SIAs move away from the dislocation. On the other hand, the zone on the right of the dislocation should attract the SIAs since their elastic energy decreases as they approach the dislocation. From Fig. \ref{fig:Eint_2D_SIA_trapped_DL} we also can distinguish the region where the dislocation loop is located, on the right of the dislocation line. We can see that the plane of the DL is a zone where the interaction energy of a SIA increases from the edges of the DL up to a maximum at its center, being thus a repulsion zone for the SIAs. This explains why the zones above and below the DL are depleted from SIAs (see Fig. \ref{fig:SIAs_trapped_DL} (Right)). These findings are in very good agreement with those recently obtained by Yu and Xu \cite{Yu2023} using MD to determine the SIA diffusion pathways near a $\langle100\rangle$ DL in Fe.

Fig. \ref{fig:Eint_2D_SIA_trapped_DL} also reveals that there are two regions located at the edge of the DL and near the dislocation line where the interaction energy of the SIA decreases. In the zones located at the edges of the loop, the interaction energy decreases down to approximately -0.20 eV whereas in the region between the loop and the dislocation, it decreases down to approximately -0.40 eV, due to the sum of the elastic fields. These potential wells near the loop are comparable to the migration energy of the SIA (0.34 eV) and explains why most of the SIAs that evolve in the surrounding of the trapped DL ends up located near the DL.
Indeed, when a SIA is attracted to this region, it becomes difficult for it to escape as it must surmount now an energy barrier of at least $0.34 + 0.20 = 0.54$ eV to migrate. This suggests that when a DL is trapped near a dislocation line, the agglomeration of SIAs to the DL could become preferential, which could enhance its growth.

\section{Conclusions}
A novel approach based on FFT solvers is proposed to include elastic interactions in OkMC simulations of defect evolution in materials, including both point defects, extended defects such as dislocation loops and dislocation lines. The method is based in the \emph{exact} determination of interaction energy by integration of the energy density map of the defect ensemble and the substraction of the energies of each individual interacting defect. The strain field of a defect is obtained solving a mechanical problem with eigenstrains, which for each defect is taken from analytical expressions or, in the case of dislocations, solving a static field dislocation mechanics. Both the mechanical equilibrium and the field dislocation mechanics problems are efficiently solved in Fourier space taking advantage of the efficiency of the FFT algorithm. Moreover, to reduce the computation time, the strain field of each type of defect is obtained once and stored in Fourier space to be translated to its actual positions during the OkMC simulation using the shift theorem. Contrary to interaction energy expressions obtained with the dipole approach, the present method does not introduce any approximation in the elastic interaction and can be used with anisotropic metals at the same computational cost.

The method is adapted to the parallel OkMC approach \cite{Jimenez2016} by taking into account the effect of the elastic fields on the effective migration barriers of defects. To determine the jump probability of a defect in all the possible directions, the spatial variation of elastic interaction energy is added to the original migration barrier of the defect. Thereby, in the presence of elastic fields, the jump probabilities of defects are biased and the direction of elastic energy decrease is the most probable one. The introduction of the \emph{exact} energy interaction in the OkMC method allows to resolve complex scenarios such as the evolution of self-interstitial atoms (SIAs) ensembles near a dislocation loop and under the presence of dislocations.

The method developed has been used to model the elastic interactions of several defect types including SIAs, dislocation loops (DLs) and straight dislocations and to predict the evolution of ensembles of these type of defects in iron. The conclusion of these results,
\begin{itemize}
    \item Interaction energies of dislocation loops with themselves or with straight dislocations computed using the current approach are equal to dipole based expressions for sufficiently far dislocations and isotropic matrix
    \item For close dislocation loops, the present method based on \emph{exact} interaction energy calculation predicts higher interaction energies than dipole approximate expressions, leading to more stable configurations 
    \item The introduction of anisotropy increases the differences of the present approach with respect to approximated expressions 

\end{itemize}

\section{Acknowledgements}
Javier Segurado and Rodrigo Santos-Güemes acknowledge the Spanish Ministry of Science for the project ADSORBENT, Plan estatal de I+D+i- 2019: PID2019-106759GB-I00.

This work has been carried out within the framework of the EUROfusion Consortium, funded by the European Union via the Euratom Research
and Training Programme (Grant Agreement No 101052200 —
EUROfusion). Views and opinions expressed are however those of the
author(s) only and do not necessarily reflect those of the European
Union or the European Commission. Neither the European Union nor the
European Commission can be held responsible for them.
One of the authors (C.J.O.) would like to thank Pui-Wai Ma and Christian Robertson for helpful discussions. C.J.O. also aknowledges Carlos Guerard for valuable discussions, which allowed him expanding his bicameral mind.

\clearpage
\appendix

\section{Dipole tensor as function of inelastic strain}\label{AppendixB}
The equivalence of the definition of a dipole tensor in terms of the integral of elastic strains in all the domain $\Omega_d$ or the integral of the inelastic strains only the domain where inelastic strain exist $\Omega_i$ is presented next. The derivation is equivalent to the elastic strain interaction in the book of Mura \cite{Mura87}.
\noindent
Let $\EPS^{d,e}=\EPS^{d}-\EPS^{i}$, then dipole expression in Eq. \ref{eq:dipole_def} can be expressed as
\begin{eqnarray}
\mathbf{P}^d =- \int_{\Omega_d} \mathbb{C}:\EPS^{d,e}(\mathbf{x}) \mathrm{d}  \Omega  =- \int_{\Omega_d} \mathbb{C}:(\EPS^{d}-\EPS^{d,i}) \mathrm{d}  \Omega=\nonumber \\ \underbrace{- \int_{\Omega_d} \mathbb{C}:\EPS^{d} \mathrm{d}  \Omega}_{\mathbf{a}} + \int_{\Omega_i} \mathbb{C}:\EPS^{d,i}(\mathbf{x}) \mathrm{d}  \Omega.
\label{eq:dip3}
\end{eqnarray}
The total strain $\EPS^{d}$ is compatible, and can be written as
\begin{equation}
\EPS^{d} = \text{grad}^s \mathbf{u}^d.
\label{eq:ugrad}
\end{equation}
Considering Eq. \eqref{eq:ugrad}, the first term of the right hand side in Eq. \eqref{eq:dip3} can be written as
$$
\mathbf{a}=\int_{\Omega_d} \mathbb{C}:\EPS^{d} \mathrm{d}  \Omega  = \int_{\Omega_d} \mathbb{C}:\text{grad}^s \mathbf{u}^{d} \mathrm{d}  \Omega 
$$
and integrating by parts
$$
 \int_{\Omega_d} \mathbb{C}:\text{grad}^s \mathbf{u}^{d} \ \mathrm{d}  \Omega = \int_{\Gamma}  \mathbb{C}:\mathbf{u}\otimes \mathbf{n}  \ \mathrm{d}  \Gamma = \mathbb{C}: \int_{\Gamma} \mathbf{u}\otimes \mathbf{n} \ \mathrm{d}  \Gamma = \mathbf{0}
 $$
 where $\Gamma$ is the surface of $\Omega_d$ and $\mathbf{n}$ its external normal. In an infinite body, the integral is zero because $\Omega_d$ is taken sufficiently large such that $\EPS$ is zero near its surface, so $\mathbf{u}$ is constant in $\Gamma$ and the integral vanishes. Note that in the case of periodicity in $\Gamma$, the integral vanishes in the cell independently on its size due to the equal of value $\mathbf{u}$ on opposite sides of the cell.
Therefore, the resulting equation is the common definition of the dipole tensor,
\begin{equation}
\mathbf{P}^d =  \int_{\Omega_i} \mathbb{C}:\EPS^{d,i}(\mathbf{x}) \mathrm{d}  \Omega = \Omega_i  \mathbb{C}:\overline{\EPS}^{d,i}.
\end{equation}

\section{Pseudocode for OkMC model with elastic interaction energies from FFT calculations}\label{AppendixA}
    \begin{algorithm}[H]
\caption{}
\label{algo}
\begin{minipage}{\textwidth}
\renewcommand\footnoterule{}  
\begin{algorithmic}[1]
 
 \State $\nu_{max} \gets \max(\nu_1,...\nu_i,..\nu_N)$ ; $\delta t \gets \frac{1}{\nu_{max}}$
 
 \item[]
 
 \ForAll{defects $d_i$}
  \State Calculate elastic strain in reference position $\boldsymbol{\EPS}^{d_0,e}_i(\mathbf{x})$\footnote{This is done only for each different type of defect}
 \EndFor
 
 \item[]

 \State Set initial position $\mathbf{X}^d_i(t=0)$ of defects $d_i$
 \ForAll{defects $d_i$}
  \State Move elastic strain fields to the current positions $\mathbf{X}^{d}_i(t=0)$ using shift Theorem, Eq. \eqref{eq:shift}: $\boldsymbol{\EPS}^{d,e}(\mathbf{X}^d_i; \mathbf{x})$
 \EndFor
 
 \State Calculate initial total elastic strain $\boldsymbol{\EPS}^{e}(\mathbf{x})=\sum_i \EPS^{d,e}(\mathbf{X}^d_i(t=0);\mathbf{x})$ 
 \item[]
 \State $t \gets 0$
 \While{$t < t_{end}$}
 
  
  \ForAll{$d_i$}
  
   \State Calculate number of events $N_i$ to perform during $\delta t$ using Eq. \eqref{eq:poisson}
   
   
   \If{$N_i > 0$}
   
    \State Subtract $\boldsymbol{\EPS}^{d,e}(\mathbf{X}^d_i(t);\mathbf{x})$  from current elastic strain field $\boldsymbol{\EPS}^{e}(\mathbf{x})$:
    \State $\boldsymbol{\EPS}'(\mathbf{x}) \gets \boldsymbol{\EPS}^e(\mathbf{x}) - \boldsymbol{\EPS}^{d,e}(\mathbf{X}^d_i(t);\mathbf{x})$
    
   
    \ForAll{possible jumps $\boldsymbol{\lambda_j}$}
    
     \State $\mathbf{X}^d_{i'} \gets \mathbf{X}^d_i(t) + \boldsymbol{\lambda_j}$
     \State Move strain field to $\mathbf{X}^d_{i'} $, $\boldsymbol{\EPS}^{d,e}(\mathbf{X}^d_{i'}\mathbf{x})$ 
     \State $\boldsymbol{\EPS}''(\mathbf{x}) \gets \boldsymbol{\EPS}'(\mathbf{x}) + \boldsymbol{\EPS}^{d,e}(\mathbf{X}^{d,e}_{i'}\mathbf{x})$      \State Calculate $E''$ corresponding to $\boldsymbol{\EPS}''(\mathbf{x})$
     \State Calculate $\Delta E''= E'' - E(t)$
     \State Calculate biased jump frequency $\nu_i^j$ using Eq. \eqref{eq:biased_forward_boltzmann}
     
    \EndFor
    
    \State Select randomly jump $j$ with $\nu_i^j$ and update position $\mathbf{X}^d_{i}(t+\delta t)$
   
   \EndIf
   
  \EndFor
  
  \State Update elastic strain $\boldsymbol{\EPS}^{e}(\mathbf{x})=\sum_i \EPS^{d,e}(\mathbf{X}^d_i(t+\delta t)  ;\mathbf{x})$ and energy $E(t+\delta t)$

  \State $t \gets t + \delta t$
 \EndWhile
\end{algorithmic}
\end{minipage}
\end{algorithm}

\bibliographystyle{unsrt}

\end{document}